\documentclass[12pt]{article}
\usepackage{amsmath,amsthm}
\usepackage{eufrak}
\usepackage{pb-diagram,lamsarrow,pb-lams}
\usepackage{latexsym}

\newcommand{\ar}{\renewcommand{\arraystretch}{1}} 
\DeclareMathAlphabet{\bb}{U}{msb}{m}{n}
\gdef\C{\bb C}

\gdef\R{\bb R}

 \DeclareMathOperator{\spin}{{\bf
Spin}}

\DeclareMathOperator{\Sym}{Sym}

 \DeclareMathOperator{\SL}{SL}
\DeclareMathOperator{\SO}{SO}\DeclareMathOperator{\SU}{SU}
\DeclareMathOperator{\Sp}{Sp}

\newcommand{\re}{\mbox{\rm Re}\,}

\newcommand{\cP}{{\cal P}}
\newcommand{\cL}{\mathcal{L}}
\newcommand{\cM}{{\cal M}}
\newcommand{\cH}{{\cal H}}

\newcommand{\sL}{\Lambda}

\newcommand{\bk}{{\bf k}}
\newcommand{\bp}{{\bf p}}
\newcommand{\bx}{{\bf x}}
\newcommand{\by}{{\bf y}}
\newcommand{\bz}{{\bf z}}
\newcommand{\bB}{{\bf B}}

\newcommand{\bE}{{\bf E}}

\newcommand{\fM}{\mathfrak{M}}
\newcommand{\fG}{\mathfrak{G}}
\newcommand{\fC}{\mathfrak{C}}

\newcommand{\fP}{\mathfrak{P}}

\newcommand{\fL}{\mathfrak{L}}

\newcommand{\fS}{\mathfrak{S}}
\newcommand{\fZ}{\mathfrak{Z}}

\newcommand{\fc}{\mathfrak{c}}
\newcommand{\fg}{\mathfrak{g}}
\newcommand{\fq}{\mathfrak{q}}

\newcommand{\Lip}{\boldsymbol{\Gamma}}

\newcommand{\balpha}{\boldsymbol{\alpha}}
\newcommand{\cl}{C\kern -0.2em \ell}

\newcommand{\e}{\mbox{\bf e}}

\newcommand{\hypergeom}[5]{\mbox{$
_#1 F_#2\left.
\!\!
\left(
\!\!
\begin{array}{c}
\multicolumn{1}{c}{\begin{array}{c}
#3
\end{array}}\\[1mm]
\multicolumn{1}{c}{\begin{array}{c}
#4
\end{array}}\end{array}
\!\!
\right|\displaystyle{#5}\right)
$}
}

\newcommand{\lf}{\left\{}
\newcommand{\rf}{\right\}}
\begin{document}
\title{Convergence of quantum electrodynamics on the Poincar\'{e} group}
\author{V.~V. Varlamov\thanks{Siberian State Industrial University,
Kirova 42, Novokuznetsk 654007, Russia}}
\date{}
\maketitle
\begin{abstract}
Extended particles are considered in terms of the fields on the
Poincar\'{e} group. Dirac like wave equations for extended particles
of any spin are defined on the various homogeneous spaces of the
Poincar\'{e} group. Free fields of the spin 1/2 and 1 (Dirac and
Maxwell fields) are considered in detail on the eight-dimensional
homogeneous space, which is equivalent to a direct product of
Minkowski spacetime and two-dimensional complex sphere. It is shown
that a massless spin-1 field, corresponding to a photon field,
should be defined within principal series representations of the
Lorentz group. Interaction between spin-1/2 and spin-1 fields is
studied in terms of a trilinear form. An analogue of the Dyson
formula for $S$-matrix is introduced on the eight-dimensional
homogeneous space. It is shown that in this case elements of the
$S$-matrix are defined by convergent integrals.
\end{abstract}
{\bf Keywords}: extended particles, fields on the Poincar\'{e}
group, homogeneous spaces, wave equations, quantum electrodynamics\\
PACS numbers:\;{\bf 02.30.Gp, 02.60.Lj, 03.65.Pm, 12.20.-m}

\section{Introduction}
It is well-known that the representation of a particle as an
idealized point has been used in physics for a long time.
Historically, such a description follows from classical (celestial)
mechanics, where the distances between explored objects are much
greater than their sizes. However, in quantum field theory a
point-like particle description meets with some conceptual
difficulties. For example, ultra-violet divergences follow directly
from the point-like description. Refusing from the point-like
idealization, we come to \emph{extended objects}.

Our consideration based on the concept of generalized wave functions
introduced by Ginzburg and Tamm in 1947 \cite{GT47}, where the wave
function depends both coordinates $x_\mu$ and additional internal
variables $u_\mu$ which describe spin of the particle,
$\mu=0,1,2,3$. In 1955, Finkelstein showed \cite{Fin55} that
elementary particles models with internal degrees of freedom can be
described on manifolds larger then Minkowski spacetime (homogeneous
spaces of the Poincar\'{e} group). The quantum field theories on the
Poincar\'{e} group were discussed in the papers
\cite{Lur64,Kih70,BF74,Aro76,KLS95,Tol96,LSS96,Dre97,GS01,GL01}. A
consideration of the field models on the homogeneous spaces leads
naturally to a generalization of the concept of wave function
(fields on the Poincar\'{e} group). The general form of these fields
relates closely with the structure of the Lorentz and Poincar\'{e}
group representations \cite{GMS,Nai58,BBTD88,GS01} and admits the
following factorization $f(x,\bz)=\phi^n(\bz)\psi_n(x)$, where $x\in
T_4$ and $\phi^n(\bz)$ form a basis in the representation space of
the Lorentz group. At this point, four parameters $x^\mu$ correspond
to position of the point-like object, whereas remaining six
parameters $\bz\in\spin_+(1,3)$ define orientation in quantum
description of orientable (extended) object \cite{GS09,GS10} (see
also \cite{Kai09}). It is obvious that the point-like object has no
orientation, therefore, orientation is an intrinsic property of the
extended object. Taking it into account, we come to consideration of
physical quantity as an extended object, the generalized wave
function of which is described by the field
\[
\boldsymbol{\psi}(\balpha)\equiv\langle
x,g\,|\boldsymbol{\psi}\rangle
\]
on the homogeneous space of some orthogonal group $\SO(p,q)$, where
$x\in T_n$ (position) and $g\in\spin_+(p,q)$ (orientation), $n=p+q$.
So, in \cite{SZ92,SZ94} Segal and Zhou proved convergence of quantum
field theory, in particular, quantum electrodynamics, on the
homogeneous space $R^1\times S^3$ of the conformal group $\SO(2,4)$,
where $S^3$ is the three-dimensional real sphere.

On the other hand, measurements in quantum field theory lead to
extended objects. As is known, loop divergences emerging in the
Green functions in quantum field theory originate from
correspondence of the Green functions to \emph{unmeasurable} (and
hence unphysical) point-like quantities. This is because no physical
quantity can be measured in a point, but in a region, the size of
which (or `diameter'\footnote{For example, this diameter can be
defined by a characteristic rest-frame kernel dimension
$\sim\;\hbar/mc$ (Compton wavelength). The magnitude of the `kernel'
radius of extended particles was first proposed by de Broglie
\cite{Bro57}, see also \cite{Tod00,Sas99}.} of the extended object)
is constrained by the resolution of measuring equipment
\cite{Alt10}. Ordinary quantum field theory defines the field
function $\phi(x)$ as a scalar product of the state vector of the
system and the state vector corresponding to the localization at the
point:
\[
\phi(x)\equiv\langle x|\phi(x)\rangle.
\]
This field describes the point-like object. On the other hand,
\emph{resolution-dependent fields}
\[
\phi_a(x)\equiv\langle x,a;g|\phi(x)\rangle,
\]
introduced in the work \cite{Alt10}, describe in essence extended
objects, here $\langle x,a;g|$ is the bra-vector corresponding to
localization of the measuring device around the point with the
spatial resolution $a$, $g$ labels the apparatus function of the
equipment (an aperture). It is easy to see that the
resolution-dependent fields have the same mathematical structure as
the fields on the Poincar\'{e} group (more generally, \emph{fields
on the groups}).

The present paper is organized as follows. Basic facts concerning
fields on the Poincar\'{e} group are considered in the section 2.
Dirac-like wave equations for extended objects are presented in the
section 3. Solutions for the fields
$\boldsymbol{\psi}(\balpha)=\langle
x,\fg\,|\boldsymbol{\psi}\rangle$ of spin 1/2 and 1 (Dirac and
Maxwell fields) are given in terms of associated hyperspherical
functions defined on the two-dimensional complex sphere, where
$\fg\in\spin_+(1,3)$. Interaction between these fields is introduced
in the section 4 in terms of a trilinear form.

\section{Fields on the Poincar\'{e} group}
As it mentioned above, Ginzburg and Tamm proposed to consider new
internal continuous variables $u_\mu$. It allows one to generalize
the Klein-Gordon wave equation. The Ginzburg-Tamm equation has the
form
\begin{equation}\label{GT1}
\left[\Box-m^2+\frac{\beta}{2}M_{\mu\nu}M_{\mu\nu}\right]\Psi(x_\rho,u_\gamma)=0,
\end{equation}
where
\[
M_{\mu\nu}=-i\left(u_\mu\frac{\partial}{\partial
u_\nu}-u_\nu\frac{\partial}{\partial
u_\mu}\right)\quad(\mu,\nu=0,1,2,3)
\]
are rotation generators of the four-dimensional space, the constant
$\beta$ is analogous to the momentum $J$ in the rotator hamiltonian
$\cH=(2J)^{-1}L_{ik}L_{ik}$, where $L_{ik}$ are generators of the
Lorentz group, $m^2$ is a constant. The wave function $\Psi$ depends
on coordinates $x_\mu$ of mass centre of the particle and internal
variables $u_\mu$. The 4-vector $u_\mu$ is a 4-vector of relative
motion of the structural component of the particle around its mass
centre. The wave function $\Psi$ can be factored:
\[
\Psi(x_\mu,u_\mu)=\Psi(x_\mu)\Phi(u_\mu).
\]
Then the Ginzburg-Tamm equation is decomposed in the following two
equations:
\begin{equation}\label{GT2}
(\Box - m^2+\lambda_1\beta)\Psi(x_\mu)=0,\quad
L_1\Phi(u_\mu)=\lambda_1\Phi(u_\mu).
\end{equation}
At this point, the function $\Phi(u_\mu)$ is an eigenfunction of the
operator $L_1$,
\[
L_1=M_{\mu\nu}M_{\mu\nu}/2,\quad L_1\Phi=\lambda_1\Phi.
\]
The first equation from (\ref{GT2}) has the following solution:
\[
\Psi(x_\mu)=C\exp(-i m_0t),
\]
where
\[
m^2_0=m^2-\lambda_1\beta.
\]
Thus, a mass spectrum, described by the equation (\ref{GT1}), is
defined by an eigenvalue of the operator $L_1$. The mass operator
$L_1$ in spherical coordinates on the one-sheeted hyperboloid $H^3$,
\[
\begin{array}{ccl}
u_0&=&r\sinh\chi;\\
u_1&=&r\cosh\chi\sin\theta\cos\phi;
\end{array}\quad
\begin{array}{ccl}
u_2&=&r\cosh\chi\sin\theta\sin\phi;\\
u_3&=&r\cosh\chi\cos\theta,
\end{array}
\]
is defined by an expression
\begin{multline}
L_1=\frac{1}{r}\frac{\partial}{\partial
r}\left(r^3\frac{\partial}{\partial
r}\right)-\frac{1}{\cosh^2\chi}\frac{\partial}{\partial\chi}\left(\cosh^2\chi\frac{\partial}{\partial\chi}\right)-\\
-\frac{1}{\cosh^2\chi}\left[\frac{1}{\sin\theta}\frac{\partial}{\partial\theta}\left(\sin\theta\frac{\partial}
{\partial\theta}\right)+\frac{1}{\sin^2\theta}\frac{\partial^2}{\partial\phi^2}\right]-
\left(r\frac{\partial}{\partial
r}\right)^2-2r\frac{\partial}{\partial r}. \nonumber
\end{multline}
\emph{The hyperboloid $H^3$ is a homogeneous space of the Lorentz
group}. Solutions of the Ginzburg-Tamm equation have the form of the
fields on the Poincar\'{e} group:
\[
\Psi(x_\mu,u_\mu)=C\exp(-i
m_0t)Y_{l,m}(\theta,\phi)P^j_l(\tanh\chi),
\]
where $Y_{l,m}(\theta,\phi)$ is a spherical function, and
$P^j_l(\tanh\chi)$ is a Legendre polynomial.

Further, fields on the Poincar\'{e} group present itself a natural
generalization of the concept of wave function. These fields
(generalized wave functions) were introduced independently by
several authors \cite{GT47,BW48,Yuk50,Shi51} mainly in connection
with constructing relativistic wave equations. In essence, this
generalization consists in replacing the Minkowski space by a larger
space on which the Poincar\'{e} group acts. If this action is to be
transitive, one is lead to consider the homogeneous spaces of the
Poincar\'{e} group. All the homogeneous spaces of this type were
listed by Finkelstein \cite{Fin55} and by Bacry and Kihlberg
\cite{BK69} and the fields on these spaces were considered in the
works \cite{Lur64,BN67,NB67,Kih68,Kih70,Tol96,GS01}.

A homogeneous space $\cM$ of a group $G$ has the following properties:\\
a) It is a topological space on which the group $G$ acts
continuously, that is, let $y$ be a point in $\cM$, then $gy$ is
defined and is again
a point in $\cM$ ($g\in G$).\\
b) This action is transitive, that is, for any two points $y_1$ and
$y_2$ in $\cM$ it is always possible to find a group element $g\in
G$
such that $y_2=gy_1$.\\
There is a one-to-one correspondence between the homogeneous spaces
of $G$ and the coset spaces of $G$. Let $H_0$ be a maximal subgroup
of $G$ which leaves the point $y_0$ invariant, $gy_0=y_0$, $g\in
H_0$, then $H_0$ is called the stabilizer of $y_0$. Representing now
any group element of $G$ in the form $g=g_cg_0$, where $g_0\in H_0$
and $g_c\in G/H_0$, we see that, by virtie of the transitivity
property, any point $y\in\cM$ can be given by $y=g_cg_0y_0=g_cy$.
Hence it follows that the elements $g_c$ of the coset space give a
parametrization of $\cM$. The mapping $\cM\leftrightarrow G/H_0$ is
continuous since the group multiplication is continuous and the
action on $\cM$ is continuous by definition. The stabilizers $H$ and
$H_0$ of two different points $y$ and $y_0$ are conjugate, since
from $H_0g_0=g_0$, $y_0=g^{-1}y$, it follows that $gH_0g^{-1}y=y$,
that is, $H=gH_0g^{-1}$.

Returning to the Poincar\'{e} group $\cP$, we see that the
enumeration of the different homogeneous spaces $\cM$ of $\cP$
amounts to an enumeration of the subgroups of $\cP$ up to a
conjugation. Following to Finkelstein, we require that $\cM$ always
contains the Minkowski space $\R^{1,3}$ which means that four
parameters of $\cM$ can be denoted by $x\,(x^\mu)$. This means that
the stabilizer $H$ of a given point in $\cM$ can never contain an
element of the translation subgroup of $\cP$. Thus, the stabilizer
must be a subgroup of the homogeneous Lorentz group $\fG_+$.

In such a way, studying different subgroups of $\fG_+$, we obtain a
full list of homogeneous spaces $\cM=\cP/H$ of the Poincar\'{e}
group. In the present paper we restrict ourselves by a consideration
of the following four homogeneous spaces:
\begin{eqnarray}
\cM_{10}&=&\R^{1,3}\times\fL_6,\quad H=0;\nonumber\\
\cM_8&=&\R^{1,3}\times S^2_c,\quad H=\Omega^c_\psi;\nonumber\\
\cM_7&=&\R^{1,3}\times H^3,\quad H=\SU(2);\nonumber\\
\cM_6&=&\R^{1,3}\times S^2,\quad
H=\{\Omega^c_\psi,\Omega_\tau,\Omega_\epsilon\}. \nonumber
\end{eqnarray}
Hence it follows that a group manifold of the Poincar\'{e} group,
$\cM_{10}=\R^{1,3}\times\fL_6$, is a maximal homogeneous space of
$\cP$, $\fL_6$ is a group manifold of the Lorentz group. The fields
on the manifold $\cM_{10}$ were considered by Lur\c{c}at
\cite{Lur64}. These fields depend on all the ten parameters of
$\cP$:
\[
\boldsymbol{\psi}(\balpha)=\langle
x,\fg\,|\boldsymbol{\psi}\rangle=\psi(x_0,x_1,x_2,x_3)
\psi(\fg_1,\fg_2,\fg_3,\fg_4,\fg_5,\fg_6),
\]
where an explicit form of $\psi(x)$ is given by the exponentials,
and the functions $\psi(\fg)$ are expressed via the generalized
hyperspherical functions $\fM^l_{mn}(\fg)$. As is known, the
universal covering of the proper Poincar\'{e} group is isomorphic to
a semidirect product $\SL(2;\C)\odot T_4$ or $\spin_+(1,3)\odot
T_4$. Since the group $T_4$ is Abelian, then all its representations
are one-dimensional. Thus, all the finite-dimensional
representations of the proper Poincar\'{e} group in essence are
equivalent to the representations $\fC$ of the group $\spin_+(1,3)$.
In the case of finite-dimensional representations of $\spin_+(1,3)$,
\begin{equation}\label{TenRep}
T_{\fg}q(\xi,\overline{\xi})=(\gamma\xi+\delta)^{l_0+l_1-1}
\overline{(\gamma\xi+\delta)}^{l_0-l_1+1}q\left(\frac{\alpha\xi+\beta}{\gamma\xi+\delta};
\frac{\overline{\alpha\xi+\beta}}{\overline{\gamma\xi+\delta}}\right),
\end{equation}
basic functions (matrix elements) of the finite-dimensional
representation of $\cP$ have the form
\[
t^l_{mn}(\boldsymbol{\alpha})=e^{-imx}\fM^l_{mn}(\fg),
\]
where
\[
\fM^l_{mn}(\fg)=e^{-im\varphi^c}\fZ^l_{mn}(\cos\theta^c)e^{-in\psi^c}.
\]
\emph{Hyperspherical functions} $\fZ^l_{mn}(\cos\theta^c)$ can be
expressed via the hypergeometric functions \cite{Var06}. So, at
$m\geq n$ and $m\geq k,\;k\geq n$ we have
\begin{multline}
\fZ^l_{mn}(\cos\theta^c)=i^{m-n}
\sqrt{\frac{\Gamma(l+m+1)\Gamma(l-n+1)}{\Gamma(l-m+1)\Gamma(l+n+1)}}
\cos^{2l}\frac{\theta}{2}\cosh^{2l}\frac{\tau}{2}\times\\
\times\sum^l_{k=-l}\tan^{m-k}\frac{\theta}{2}\tanh^{k-n}\frac{\tau}{2}\times\\
\times\hypergeom{2}{1}{m-l,-k-l}{m-k+1}{-\tan^2\frac{\theta}{2}}
\hypergeom{2}{1}{k-l,-n-l}{k-n+1}{\tanh^2\frac{\tau}{2}}.
\label{HSF}
\end{multline}
There also exist three expressions of the hypergeometric type for
the functions $\fZ^l_{mn}(\cos\theta^c)$ with the index values
$m\geq n,\;m\geq k,\;n\geq k$, $n\geq m,\;k\geq m,\;n\geq k$ and
$n\geq m,\;k\geq m,\;k\geq n$. For the unitary representations, that
is, in the case of principal series representations of the group
$\SO_0(1,3)$, there exists an analogue of the formula
(\ref{TenRep}),
\begin{equation}\label{Principal}
V_af(z)=(a_{12}z+a_{22})^{\frac{\lambda}{2}+i\frac{\rho}{2}-1}
\overline{(a_{12}z+a_{22})}^{-\frac{\lambda}{2}+i\frac{\rho}{2}-1}
f\left(\frac{a_{11}z+a_{21}}{a_{12}z+a_{22}}\right),
\end{equation}
where $f(z)$ is a measurable function of the Hilbert space $L_2(Z)$,
satisfying the condition $\int|f(z)|^2dz<\infty$, $z=x+iy$. A
totality of all representations $a\rightarrow T^\alpha$,
corresponding to all possible pairs $\lambda$, $\rho$, is called a
principal series of representations of the group $\SO_0(1,3)$ and
denoted as $\fS_{\lambda,\rho}$. At this point, a comparison of
(\ref{Principal}) with the formula (\ref{TenRep}) for the spinor
representation $\fS_l$ shows that the both formulas have the same
structure; only the exponents at the factors $(a_{12}z+a_{22})$,
$\overline{(a_{12}z+a_{22})}$ and the functions $f(z)$ are
different. In the case of spinor representations the functions
$f(z)$ are polynomials $p(z,\bar{z})$ in the spaces $\Sym_{(k,r)}$,
and in the case of a representation $\fS_{\lambda,\rho}$ of the
principal series $f(z)$ are functions from the Hilbert space
$L_2(Z)$. Further, we know that a representation $S_l$ of the group
$\SU(2)$ is realized in terms of the functions
$t^l_{mn}(u)=e^{-im\varphi}P^l_{mn}(\cos\theta)e^{-in\psi}$
\cite{Vil68}. It is well known that the representation $S_k$ of
$\SU(2)$ is contained in $\fS_{\lambda,\rho}$ not more then one time
\cite{Nai58}. At this point, $S_k$ is contained in
$\fS_{\lambda,\rho}$, when $\frac{\lambda}{2}$ is one from the
numbers $-k,-k+1,\ldots,k$.
Matrix elements of the principal series representations of the group
$\SO_0(1,3)$ have the form \cite{Var06}
\begin{multline}
\fM^{-\frac{1}{2}+i\rho,l_0}_{mn}(\mathfrak{g})=
e^{-m(\epsilon+i\varphi)-n(\varepsilon+i\psi)}
\fZ^{-\frac{1}{2}+i\rho,l_0}_{mn}=
e^{-m(\epsilon+i\varphi)-n(\varepsilon+i\psi)}\times\\[0.2cm]
\sum^{l_0}_{t=-l_0}i^{m-t} \sqrt{\Gamma(l_0-m+1)
\Gamma(l_0+m+1)\Gamma(l_0-t+1)
\Gamma(l_0+t+1)}\times\\
\cos^{2l_0}\frac{\theta}{2}\tan^{m-t}\frac{\theta}{2}\times\\[0.2cm]
\sum^{\min(l_0-m,l_0+t)}_{j=\max(0,t-m)}
\frac{i^{2j}\tan^{2j}\frac{\theta}{2}} {\Gamma(j+1)\Gamma(l_0-m-j+1)
\Gamma(l_0+t-j+1)\Gamma(m-t+j+1)}\times\\[0.2cm]
\sqrt{\Gamma(\tfrac{1}{2}+i\rho-n)
\Gamma(\tfrac{1}{2}+i\rho+n)\Gamma(\tfrac{1}{2}+i\rho-t)
\Gamma(\tfrac{1}{2}+i\rho+t)}
\cosh^{-1+2i\rho}\frac{\tau}{2}\tanh^{n-t}\frac{\tau}{2}\times\\[0.2cm]
\sum^{\infty}_{s=\max(0,t-n)} \frac{\tanh^{2s}\frac{\tau}{2}}
{\Gamma(s+1)\Gamma(\frac{1}{2}+i\rho-n-s)
\Gamma(\frac{1}{2}+i\rho+t-s)\Gamma(n-t+s+1)},\label{MPrincip}
\end{multline}
where $l_0=\left|\frac{\lambda}{2}\right|$ and $\frac{\lambda}{2}$
is one from the numbers $-k,-k+1,\ldots,k$. It is obvious that
$\fM^{-\frac{1}{2}+i\rho,l_0}_{mn}(\fg)$ cannot be attributed as
matrix elements to single irreducible representation. From the
latter expression it follows that relativistic spherical functions
$f(\fg)$ of the principal series can be defined by means of the
function
\begin{equation}\label{MEPS}
\fM^{-\frac{1}{2}+i\rho,l_0}_{mn}(\fg)=e^{-m(\epsilon+i\varphi)}
\fZ^{-\frac{1}{2}+i\rho,l_0}_{mn}(\cos\theta^c)e^{-n(\varepsilon+i\psi)},
\end{equation}
where
\[
\fZ^{-\frac{1}{2}+i\rho,l_0}_{mn}(\cos\theta^c)= \sum^{l_0}_{t=-l_0}
P^{l_0}_{mt}(\cos\theta)\fP^{-\frac{1}{2}+i\rho}_{tn}(\cosh\tau).
\]
In turn, the functions
$\fM^{-\frac{1}{2}+i\rho,l_0}_{mn}(\mathfrak{g})$ can be expressed
via the hypergeometric functions. So, at $m\geq t,\;t\geq n$ we have
\cite{Var06}
\begin{multline}
\fM^{-\frac{1}{2}+i\rho,l_0}_{mn}(\fg)=i^{m-n}e^{-m(\epsilon+i\varphi)-n(\varepsilon+i\psi)}
\sqrt{\frac{\Gamma(l_0+m+1)\Gamma(i\rho-n+\frac{1}{2})}
{\Gamma(l_0-m+1)\Gamma(i\rho+n+\frac{1}{2})}}\times\\
\cos^{2l_0}\frac{\theta}{2}\cosh^{-1+2i\rho}\frac{\tau}{2}
\sum^{l_0}_{t=-l_0}
\tan^{m-t}\frac{\theta}{2}\tanh^{t-n}\frac{\tau}{2}\times\\
\hypergeom{2}{1}{m-l_0,-t-l_0}{m-t+1}{-\tan^2\frac{\theta}{2}}
\hypergeom{2}{1}{t-i\rho+\frac{1}{2},-n-i\rho+\frac{1}{2}}{t-n+1}{\tanh^2\frac{\tau}{2}}.
\label{HSU}
\end{multline}
There also exist three expressions of the hypergeometric type for
the functions $\fM^{-\frac{1}{2}+i\rho,l_0}_{mn}(\fg)$ with the
index values $m\geq t,\;n\geq t$, $t\geq m,\;n\geq t$ and $t\geq
m,\;t\geq n$.

The following eight-dimensional homogeneous space
$\cM_8=\R^{1,3}\times S^2_c$ is a direct product of the Minkowski
space $\R^{1,3}$ and the complex two-sphere $S^2_c$. In this case
the stabilizer $H$ consists of the subgroup $\Omega^c_\psi$ of the
diagonal matrices $\begin{pmatrix}
e^{\frac{i\psi^c}{2}} & 0\\
0 & e^{-\frac{i\psi^c}{2}}
\end{pmatrix}$. Bacry and Kihlberg \cite{BK69} claimed that the space
$\cM_8$ is the most suitable for a description of both half-integer
and integer spins. The fields, defined in $\cM_8$, depend on the
eight parameters of $\cP$:
\begin{equation}\label{WF}
\langle
x,\varphi^c,\theta^c|\boldsymbol{\psi}\rangle=\psi(x)\psi(\varphi^c,\theta^c)=
\psi(x_0,x_1,x_2,x_3)\psi(\varphi,\epsilon,\theta,\tau),
\end{equation}
where the functions $\psi(\varphi^c,\theta^c)$ are expressed via the
associated hyperspherical functions defined on the surface of the
complex two-sphere $S^2_c$. The sphere $S^c_2$ can be constructed
from the quantities $z_k=x_k+iy_k$, $\overset{\ast}{z}_k=x_k-iy_k$
$(k=1,2,3)$ as follows:
\begin{equation}\label{CS}
S^c_2:\;z^2_1+z^2_2+z^2_3=\bx^2-\by^2+2i\bx\by=r^2.
\end{equation}
The complex conjugate (dual) sphere $\dot{S}^c_2$ is
\begin{equation}\label{DS}
\dot{S}^c_2:\;\overset{\ast}{z}_1{}^2+\overset{\ast}{z}_2{}^2+
\overset{\ast}{z}_3{}^2=\bx^2-\by^2-2i\bx\by=\overset{\ast}{r}{}^2.
\end{equation}
For more details about the two-dimensional complex sphere see
\cite{Hus70,HS70,SH70}. The surface of $S^c_2$ is homeomorphic to
the space of the pairs $(z_1,z_2)$, where $z_1$ and $z_2$ are the
points of a \emph{complex projective line}, $z_1\neq z_2$. This
space is a homogeneous space of the Lorentz group \cite{GGV62}. In
Euler parametrization we have
$(z_1,z_2)\sim(\varphi^c,\theta^c)=(\varphi-i\epsilon,\theta-i\tau)$.
It is well-known that both quantities $\bx^2-\by^2$, $\bx\by$ are
invariant with respect to the Lorentz transformations, since a
surface of the complex sphere is invariant (Casimir operators of the
Lorentz group are constructed from such quantities). Moreover, since
the real and imaginary parts of the complex two-sphere transform
like the electric and magnetic fields, respectively, the invariance
of $\bz^2\sim(\bE+i\bB)^2$ under proper Lorentz transformations is
evident. At this point, the quantities $\bx^2-\by^2$, $\bx\by$ are
similar to the well known electromagnetic invariants $E^2-B^2$,
$\bE\bB$. This intriguing relationship between the Laplace-Beltrami
operators, Casimir operators of the Lorentz group and
electromagnetic invariants $E^2-B^2\sim\bx^2-\by^2$,
$\bE\bB\sim\bx\by$ leads naturally to a Riemann-Silberstein
representation of the electromagnetic field (see, for example,
\cite{Web01,Sil07,Bir96}). Further, associated hyperspherical
functions
\begin{equation}\label{Associated}
\fM^m_l(\varphi^c,\theta^c,0)=e^{-im\varphi^c}\fZ^m_l(\cos\theta^c),
\end{equation}
where
\[
\fZ^m_l(\cos\theta^c)=\sum^l_{k=-l}P^l_{mk}(\cos\theta)\fP^k_l(\cosh\tau),
\]
are defined on the surface of the two-dimensional complex sphere
(\ref{CS}). In turn, the functions
$\fM^{\dot{m}}_{\dot{l}}(\dot{\varphi}^c,\theta^c,0)=e^{i\dot{m}\dot{\varphi}^c}
\fZ^{\dot{m}}_{\dot{l}}(\cos\dot{\theta}^c)$ are defined on the
surface of the dual sphere (\ref{DS}). Explicit expressions of the
hypergeometric type for the functions $\fZ^m_l(\cos\theta^c)$ and
$\fZ^{\dot{m}}_{\dot{l}}(\cos\dot{\theta}^c)$ follow directly from
the previous expressions (\ref{HSF}) and (\ref{HSU}) at $n=0$. In
the case of the principal series representations of the group
$\SO_0(1,3)$ we have
\begin{equation}\label{AHPS}
\fZ^m_{-\frac{1}{2}+i\rho,l_0}(\cos\theta^c)= \sum^{l_0}_{t=-l_0}
P^{l_0}_{mt}(\cos\theta)\fP^t_{-\frac{1}{2}+i\rho}(\cosh\tau),
\end{equation}
where $\fP^t_{-\frac{1}{2}+i\rho}(\cosh\tau)$ are {\it conical
functions} (see \cite{Bat}).

In turn, a seven-dimensional homogeneous space $\cM_7=\R^{1,3}\times
H^3$ is a direct product of $\R^{1,3}$ and a three-dimensional
timelike (two-sheeted) hyperboloid $H^3$. The stabilizer $H$
consists of the subgroup of three-dimensional rotations, $\SU(2)$.
Quantum field theory on the space $\cM_7$ was studied by Boyer and
Fleming \cite{BF74}. They showed that the fields built over $\cM_7$
are in general nonlocal and become local only when the finite
dimensional representations of the Lorentz group are used. It is
easy to see that the fields $\boldsymbol{\psi}\in\cM_7$ depend on
the seven parameters of the Poincar\'{e} group:
\[
\langle
x,\tau,\epsilon,\varepsilon|\boldsymbol{\psi}\rangle=\psi(x_0,x_1,x_2,x_3)
\psi(\tau,\epsilon,\varepsilon),
\]
where the functions $\psi(\tau,\epsilon,\varepsilon)$ are expressed
via $e^{-(m\epsilon+n\varepsilon)}\fP^l_{mn}(\cosh\tau)$ in the case
of finite dimensional representations, and also via
$e^{-(m\epsilon+n\varepsilon)}\fP^{-\frac{1}{2}+i\rho}_{mn}(\cosh\tau)$
in the case of principal series of unitary representations, and via
$e^{-(m\epsilon+n\varepsilon)}\fP^{l,\pm}_{mn}(\cosh\tau)$ in the
case of the discrete representation series of the subgroup
$\SU(1,1)$. The spherical function $\fP^l_{mn}(\cosh\tau)$ on the
group $\SU(1,1)$ has the form
\begin{multline}
\fP^l_{mn}(\cosh\tau)=\sqrt{\Gamma(l-n+1)\Gamma(l+n+1)\Gamma(l-m+1)
\Gamma(l+m+1)}\times\\
\cosh^{2l}\frac{\tau}{2}\tanh^{n-m}\frac{\tau}{2}\times\\
\sum^{\min(l-n,l+m)}_{s=\max(0,m-n)}
\frac{\tanh^{2s}\frac{\tau}{2}}{\Gamma(s+1)\Gamma(l-n-s+1)\Gamma(n-m+s+1)
\Gamma(l+m-s+1)}.\label{Jacobi}
\end{multline}

Further, a six-dimensional space $\cM_6=\R^{1,3}\times S^2$ is a
minimal homogeneous space of the Poincar\'{e} group, since the real
two-sphere $S^2$ has a minimal possible dimension among the
homogeneous spaces of the Lorentz group. In this case, the
stabilizer $H$ consists of the subgroup $\Omega^c_\psi$ and the
subgroups $\Omega_\tau$ and $\Omega_\epsilon$ formed by the matrices
$\begin{pmatrix}
\cosh\frac{\tau}{2} & \sinh\frac{\tau}{2}\\
\sinh\frac{\tau}{2} & \cosh\frac{\tau}{2}
\end{pmatrix}$ and
$\begin{pmatrix}
e^{\frac{\epsilon}{2}} & 0\\
0 & e^{-\frac{\epsilon}{2}}
\end{pmatrix}$, respectively. Field models on the
homogeneous space $\cM_6$ have been considered in the works
\cite{KLS95,LSS96,Dre97}. In the paper \cite{Dre97} Drechsler
considered the real two-sphere as a `spin shell' $S^2_{r=2s}$ of
radius $r=2s$, where $s=0,\frac{1}{2},1,\frac{3}{2},\ldots$. The
fields, defined on $\cM_6$, depend on the six parameters of $\cP$:
\[
\langle
x,\varphi,\theta|\boldsymbol{\psi}\rangle=\psi(x_0,x_1,x_2,x_3)
\psi(\varphi,\theta),
\]
where the functions $\psi(\varphi,\theta)$ are expressed via the
generalized spherical functions of the type
$e^{-im\varphi}P^l_{m0}(\cos\theta)$ or via the Wigner
$D$-functions, here
\begin{multline}
P^l_{mn}(\cos\theta)=e^{-i(m\varphi+n\psi)}i^{m-n}\sqrt{\Gamma(l-m+1)\Gamma(l+m+1)\Gamma(l-n+1)
\Gamma(l+n+1)}\times\\
\cos^{2l}\frac{\theta}{2}\tan^{m-n}\frac{\theta}{2}\times\\
\sum^{\min(l-n,l+n)}_{j=\max(0,n-m)}
\frac{i^{2j}\tan^{2j}\frac{\theta}{2}}
{\Gamma(j+1)\Gamma(l-m-j+1)\Gamma(l+n-j+1)\Gamma(m-n+j+1)}\label{Mat1}
\end{multline}
is a spherical function on the group $\SU(2)$.

Further, let $\cH$ be a Hilbert space and let $U(g)$ be a square
integrable representation of a locally compact Lie group $G$ acting
transitively on $\cH$, $\forall\phi\in\cH$, $g\in
G:\,U(g)\phi\in\cH$. Let there exist such a vector
$\mid\psi\rangle\in\cH$ that satisfies the admissibility condition:
\[
C_\psi=||\psi||^{-2}\int_{g\in G}|\langle\mid
U(g)\psi\rangle|^2d\mu(g)\;<\;\infty,
\]
where $d\mu(g)$ is the left-invariant Haar measure on $G$. Then, any
vector $|\phi\rangle\in\cH$ can be represented in the form
\[
|\phi\rangle=C^{-1}_\psi\int_G\mid U(g)\psi\rangle
d\mu(g)\langle\psi U^\ast(g)\mid\phi\rangle.
\]
At the restriction of $G$ on the affine subgroup $x^\prime=ax+b$
(the group of translations and dilatations of the real axis) a
harmonic analysis on the group $G$ is reduced to the
wavelet-transformation:
\begin{eqnarray}
\phi(x)&=&\frac{1}{C_\psi}\int\frac{1}{a^d}\psi\left(\frac{x-b}{a}\right)\phi_a(b)\frac{dad^db}{a},\nonumber\\
\phi_a(b)&=&\int\frac{1}{a^d}\overline{\psi\left(\frac{x-b}{a}\right)}\phi(x)d^dx,\nonumber\\
C_\psi&=&\int^\infty_0|\widetilde{\psi}(ak)|^2\frac{da}{a}\;<\;\infty.\label{Wavelet}
\end{eqnarray}
At this restriction the field $\boldsymbol{\psi}(\balpha)=\langle
x,g\,|\boldsymbol{\psi}\rangle$ on the group $G$ is reduced to the
field $\boldsymbol{\psi}(\balpha)=\langle
x,a\,|\boldsymbol{\psi}\rangle\sim\langle x,a;g|\phi(x)\rangle$ on
the affine group, where $a$ is a scale factor which can be
associated with the resolution of measuring equipment \cite{Alt10}.
It should be noted that further restriction of $G$ on the
translation subgroup leads to the restriction of
$\boldsymbol{\psi}(\balpha)=\langle x,g\,|\boldsymbol{\psi}\rangle$
to the field $\boldsymbol{\psi}(x)=\langle
x\,|\boldsymbol{\psi}\rangle$ describing the point-like object, and
the wavelet-transformation in this case is reduced to the
Fourier-transformation. Thus, the affine group is a minimal group
which can be used for description of the extended object. The
restriction of $G$ to the affine subgroup is equivalent formally to
the Faddeev-Popov method considering translation and dilatation
invariance in quantum field calculations related with the
renormalization group, at this point, $a=1/\Lambda$, where $\Lambda$
is a cut-off parameter. Moreover, in the framework of
wavelet-analysis, loop Green functions emerging in $\phi^4$-model,
are free from ultra-violet divergences \cite{Alt07,Alt10}.

\section{Free fields on the two-dimensional complex sphere}
We will start with a more general homogeneous space of the group
$\cP$, $\cM_{10}=\R^{1,3}\times\fL_6$ (group manifold of the
Poincar\'{e} group). Let $\cL(\balpha)$ be a Lagrangian on the group
manifold $\cM_{10}$ (in other words, $\cL(\balpha)$ is a
10-dimensional point function), where $\balpha$ is the parameter set
of this group. Then an integral for $\cL(\balpha)$ on some
10-dimensional volume $\Omega$ of the group manifold we will call
{\it an action on the Poincar\'{e} group}:
\[
A=\int\limits_\Omega d\balpha\cL(\balpha),
\]
where $d\balpha$ is a Haar measure on the group $\cP$.

Let $\boldsymbol{\psi}(\balpha)=\langle
x,\fg\,|\boldsymbol{\psi}\rangle$ be a function on the group
manifold $\cM_{10}$ (now it is sufficient to assume that
$\boldsymbol{\psi}(\balpha)$ is a square integrable function on the
Poincar\'{e} group) and let
\begin{equation}\label{ELE}
\frac{\partial\cL}{\partial\boldsymbol{\psi}}-\frac{\partial}{\partial\balpha}
\frac{\partial\cL}{\partial\frac{\partial\boldsymbol{\psi}}{\partial\balpha}}=0
\end{equation}
be Euler-Lagrange equations on $\cM_{10}$ (more precisely speaking,
the equations (\ref{ELE}) act on the tangent bundle
$T\cM_{10}=\underset{\balpha\in\cM_{10}}{\cup}T_{\balpha}\cM_{10}$
of the manifold $\cM_{10}$, see \cite{Arn}). Let us introduce a
Lagrangian $\cL(\balpha)$ depending on the field function
$\boldsymbol{\psi}(\balpha)$ as follows
\[
\cL(\balpha)=-\frac{1}{2}\left(\boldsymbol{\psi}^\ast(\balpha)B_\mu
\frac{\partial\boldsymbol{\psi}(\balpha)}{\partial\balpha_\mu}-
\frac{\partial\boldsymbol{\psi}^\ast(\balpha)}
{\partial\balpha_\mu}B_\mu\boldsymbol{\psi}(\balpha)\right)
-\kappa\boldsymbol{\psi}^\ast(\balpha)B_{11}\boldsymbol{\psi}(\balpha),
\]
where $B_\nu$ ($\nu=1,2,\ldots,10$) are square matrices. The number
of rows and columns in these matrices is equal to the number of
components of $\boldsymbol{\psi}(\balpha)$, $\kappa$ is a non-null
real constant.

Further, if $B_{11}$ is non-singular, then we can introduce the
matrices
\[
\Gamma_\mu=B^{-1}_{11}B_\mu,\quad \mu=1,2,\ldots,10,
\]
and represent the Lagrangian $\cL(\balpha)$ in the form
\begin{equation}\label{Lagrange}
\cL(\balpha)=-\frac{1}{2}\left(\overline{\boldsymbol{\psi}}(\balpha)\Gamma_\mu
\frac{\partial\boldsymbol{\psi}(\balpha)}{\partial\balpha_\mu}-
\frac{\overline{\boldsymbol{\psi}}(\balpha)}{\partial\balpha_\mu}\Gamma_\mu
\boldsymbol{\psi}(\balpha)\right)-
\kappa\overline{\boldsymbol{\psi}}(\balpha)\boldsymbol{\psi}(\balpha),
\end{equation}
where
\[
\overline{\boldsymbol{\psi}}(\balpha)=\boldsymbol{\psi}^\ast(\balpha)B_{11}.
\]

Varying independently $\psi(x)$ and $\overline{\psi}(x)$, we obtain
from (\ref{Lagrange}) in accordance with (\ref{ELE}) the following
equations:
\begin{equation}\label{FET}
\begin{array}{ccc}
\Gamma_i\dfrac{\partial\psi(x)}{\partial x_i}+\kappa\psi(x)&=&0,\\
\Gamma^T_i\dfrac{\partial\overline{\psi}(x)}{\partial x_i}-
\kappa\overline{\psi}(x)&=&0.
\end{array}\quad(i=0,\ldots,3)
\end{equation}
The matrix $\Gamma_0$ in equations (\ref{FET}) can be written in the
form (see \cite{GMS,AD72,PS83})
\begin{equation}\label{G1}
\Gamma_0=\text{diag}\left(C^0\otimes I_1,C^1\otimes I_3,\ldots,
C^s\otimes I_{2s+1},\ldots\right)
\end{equation}
for integer spin and
\begin{equation}\label{G2}
\Gamma_0=\text{diag}\left(C^{\frac{1}{2}}\otimes I_2,
C^{\frac{3}{2}}\otimes I_4,\ldots,C^s\otimes I_{2s+1},\ldots\right)
\end{equation}
for half-integer spin, where $C^s$ is a spin block. If the spin
block $C^s$ has non-null roots, then the particle possesses the spin
$s$. The spin block $C^s$ in (\ref{G1})--(\ref{G2}) consists of the
elements $c^s_{\boldsymbol{\tau}\boldsymbol{\tau}^\prime}$, where
$\boldsymbol{\tau}_{l_1,l_2}$ and
$\boldsymbol{\tau}_{l^\prime_1,l^\prime_2}$ are interlocking
irreducible representations of the Lorentz group, that is, such
representations, for which $l^\prime_1=l_1\pm\frac{1}{2}$,
$l^\prime_2=l_2\pm\frac{1}{2}$. At this point, the block $C^s$
contains only the elements
$c^s_{\boldsymbol{\tau}\boldsymbol{\tau}^\prime}$ corresponding to
such interlocking representations $\boldsymbol{\tau}_{l_1,l_2}$,
$\boldsymbol{\tau}_{l^\prime_1,l^\prime_2}$ which satisfy the
conditions
\[
|l_1-l_2|\leq s\leq l_1+l_2,\quad |l^\prime_1-l^\prime_2|\leq s \leq
l^\prime_1+l^\prime_2.
\]
The two most full schemes of the interlocking irreducible
representations of the Lorentz group for integer and half-integer
spins are shown on the Fig.\,1 and Fig.\,2.
\begin{figure}[ht]
\[
\dgARROWLENGTH=0.5em \dgHORIZPAD=1.7em \dgVERTPAD=2.2ex
\begin{diagram}
\node[5]{(s,0)}\arrow{e,-}\arrow{s,-}\node{\cdots}\\
\node[5]{\vdots}\arrow{s,-}\\
\node[3]{(2,0)}\arrow{e,-}\arrow{s,-}\node{\cdots}\arrow{e,-}
\node{\left(\frac{s+2}{2},\frac{s-2}{2}\right)}\arrow{s,-}\arrow{e,-}
\node{\cdots}\\
\node[2]{(1,0)}\arrow{s,-}\arrow{e,-}
\node{\left(\frac{3}{2},\frac{1}{2}\right)}\arrow{s,-}\arrow{e,-}
\node{\cdots}\arrow{e,-}
\node{\left(\frac{s+1}{2},\frac{s-1}{2}\right)}\arrow{s,-}\arrow{e,-}
\node{\cdots}\\
\node{(0,0)}\arrow{e,-}
\node{\left(\frac{1}{2},\frac{1}{2}\right)}\arrow{s,-}\arrow{e,-}
\node{(1,1)}\arrow{s,-}\arrow{e,-}\node{\cdots}\arrow{e,-}
\node{\left(\frac{s}{2},\frac{s}{2}\right)}\arrow{s,-}\arrow{e,-}
\node{\cdots}\\
\node[2]{(0,1)}\arrow{e,-}
\node{\left(\frac{1}{2},\frac{3}{2}\right)}\arrow{s,-}\arrow{e,-}
\node{\cdots}\arrow{e,-}
\node{\left(\frac{s-1}{2},\frac{s+1}{2}\right)}\arrow{s,-}\arrow{e,-}
\node{\cdots}\\
\node[3]{(0,2)}\arrow{e,-}\node{\cdots}\arrow{e,-}
\node{\left(\frac{s-2}{2},\frac{s+2}{2}\right)}\arrow{s,-}\arrow{e,-}
\node{\cdots}\\
\node[5]{\vdots}\arrow{s,-}\\
\node[5]{(0,s)}\arrow{e,-}\node{\cdots}
\end{diagram}
\]
\begin{center}{\small {\bf Fig.\,1:} Interlocking representation scheme for the fields of integer spin
(Bose-scheme).}\end{center}
\end{figure}
\begin{figure}[ht]
\[
\dgARROWLENGTH=0.5em
\dgHORIZPAD=1.7em 
\dgVERTPAD=2.2ex 
\begin{diagram}
\node[4]{(s,0)}\arrow{s,-}\arrow{e,-}\node{\cdots}\\
\node[4]{\vdots}\arrow{s,-}\\
\node[2]{\left(\frac{3}{2},0\right)}\arrow{s,-}\arrow{e,-}
\node{\cdots}\arrow{e,-}
\node{\left(\frac{2s+3}{4},\frac{2s-3}{4}\right)}\arrow{s,-}\arrow{e,-}
\node{\cdots}\\
\node{\left(\frac{1}{2},0\right)}\arrow{s,-}\arrow{e,-}
\node{\left(1,\frac{1}{2}\right)}\arrow{s,-}\arrow{e,-}
\node{\cdots}\arrow{e,-}
\node{\left(\frac{2s+1}{4},\frac{2s-1}{4}\right)}\arrow{s,-}\arrow{e,-}
\node{\cdots}\\
\node{\left(0,\frac{1}{2}\right)}\arrow{e,-}
\node{\left(\frac{1}{2},1\right)}\arrow{s,-}\arrow{e,-}
\node{\cdots}\arrow{e,-}
\node{\left(\frac{2s-1}{4},\frac{2s+1}{4}\right)}\arrow{s,-}\arrow{e,-}
\node{\cdots}\\
\node[2]{\left(0,\frac{3}{2}\right)}\arrow{e,-}
\node{\cdots}\arrow{e,-}
\node{\left(\frac{2s-3}{4},\frac{2s+3}{4}\right)}\arrow{s,-}\arrow{e,-}
\node{\cdots}\\
\node[4]{\vdots}\arrow{s,-}\\
\node[4]{(0,s)}\arrow{e,-}\node{\cdots}
\end{diagram}
\]
\begin{center}{\small {\bf Fig.\,2:} Interlocking representation scheme for the fields of half-integer spin
(Fermi-scheme).}\end{center}
\end{figure}
As follows from Fig.\,1 the simplest field is the scalar field
\[
(0,0).
\]
This field is described by the Fock-Klein-Gordon equation. In its
turn, the simplest field from the Fermi-scheme (Fig.\,2) is the
electron-positron (spinor) field corresponding to the following
interlocking scheme:
\[
\dgARROWLENGTH=2.5em
\begin{diagram}
\node{\left(\frac{1}{2},0\right)}\arrow{e,<>}
\node{\left(0,\frac{1}{2}\right)}
\end{diagram}.
\]
\begin{sloppypar}\noindent
This field is described by the Dirac equation. Further, the next
field from the Bose-scheme (Fig.\,1) is a photon field (Maxwell
field) defined within the interlocking scheme\end{sloppypar}
\[
\dgARROWLENGTH=2.5em
\begin{diagram}
\node{(1,0)}\arrow{e,<>}\node{\left(\frac{1}{2},\frac{1}{2}\right)}
\arrow{e,<>}\node{(0,1)}
\end{diagram}.
\]
This interlocking scheme leads to the Maxwell equations. The fields
$(1/2,0)\oplus(0,1/2)$ and $(1,0)\oplus(0,1)$ (Dirac and Maxwell
fields) are particular cases of fields
$\boldsymbol{\psi}(\balpha)=\langle
x,\fg\,|\boldsymbol{\psi}\rangle$ of the type $(l,0)\oplus(0,l)$,
where $\fg\in\spin_+(1,3)$. Wave equations for such fields and their
general solutions were found in the works
\cite{Var03,Var04e,Var05b}.

It is easy to see that the interlocking scheme, corresponding to the
Maxwell field, contains the field of tensor type:
\[
\left(\frac{1}{2},\frac{1}{2}\right).
\]
Further, the next interlocking scheme (see Fig.\,2)
\[
\dgARROWLENGTH=2.5em
\begin{diagram}
\node{\left(\frac{3}{2},0\right)}\arrow{e,<>}\node{\left(1,\frac{1}{2}\right)}
\arrow{e,<>}\node{\left(\frac{1}{2},1\right)}\arrow{e,<>}\node{\left(0,\frac{3}{2}\right)}
\end{diagram},
\]
corresponding to the Pauli-Fierz equations \cite{FP39}, contains a
chain of the type
\[
\dgARROWLENGTH=2.5em
\begin{diagram}
\node{\left(1,\frac{1}{2}\right)}\arrow{e,<>}
\node{\left(\frac{1}{2},1\right)}
\end{diagram}.
\]
In such a way we come to wave equations for the fields
$\boldsymbol{\psi}(\balpha)=\langle
x,\fg\,|\boldsymbol{\psi}\rangle$ of tensor type
$(l_1,l_2)\oplus(l_2,l_1)$. Wave equations for such fields and their
general solutions were found in the work \cite{Var07b}.

Further, varying independently $\psi(\fg)$ and
$\overline{\psi}(\fg)$ one gets from (\ref{Lagrange}) the following
equations:
\begin{equation}\label{FEL}
\begin{array}{ccc}
\Gamma_k\dfrac{\partial\psi(\fg)}{\partial\fg_k}+\kappa\psi(\fg)&=&0,\\
\Gamma^T_k\dfrac{\overline{\psi}(\fg)}{\partial\fg_k}-
\kappa\overline{\psi}(\fg)&=&0,
\end{array}\quad(k=1,\ldots,6)
\end{equation}
where
\[
\psi(\fq)=\ar\begin{pmatrix}
\psi(\fq)\\
\dot{\psi}(\fq)
\end{pmatrix},\quad
\Gamma_k=\begin{pmatrix}
0 & \Lambda^\ast_k\\
\Lambda_k & 0
\end{pmatrix}.
\]
Non-zero elements of the matrices $\Lambda_k$ and $\Lambda^\ast_k$
have the form (for more details see \cite{Var03})
\begin{equation}\label{L1}
{\renewcommand{\arraystretch}{1.25}
\sL_1:\quad\left\{\begin{array}{ccc} a^{k^\prime k}_{l-1,l,m-1,m}&=&
-\frac{c_{l-1,l}}{2}\sqrt{(l+m)(l+m-1)},\\
a^{k^\prime k}_{l,l,m-1,m}&=&
\frac{c_{ll}}{2}\sqrt{(l+m)(l-m+1)},\\
a^{k^\prime k}_{l+1,l,m-1,m}&=&
\frac{c_{l+1,l}}{2}\sqrt{(l-m+1)(l-m+2)},\\
a^{k^\prime k}_{l-1,l,m+1,m}&=&
\frac{c_{l-1,l}}{2}\sqrt{(l-m)(l-m-1)},\\
a^{k^\prime k}_{l,l,m+1,m}&=&
\frac{c_{ll}}{2}\sqrt{(l+m+1)(l-m)},\\
a^{k^\prime k}_{l+1,l,m+1,m}&=&
-\frac{c_{l+1,l}}{2}\sqrt{(l+m+1)(l+m+2)}.
\end{array}\right.}
\end{equation}
\begin{equation}\label{L2}
{\renewcommand{\arraystretch}{1.25}
\sL_2:\quad\left\{\begin{array}{ccc} b^{k^\prime k}_{l-1,l,m-1,m}&=&
-\frac{ic_{l-1,l}}{2}\sqrt{(l+m)(l+m-1)},\\
b^{k^\prime k}_{l,l,m-1,m}&=&
\frac{ic_{ll}}{2}\sqrt{(l+m)(l-m+1)},\\
b^{k^\prime k}_{l+1,l,m-1,m}&=&
\frac{ic_{l+1,l}}{2}\sqrt{(l-m+1)(l-m+2)},\\
b^{k^\prime k}_{l-1,l,m+1,m}&=&
-\frac{ic_{l-1,l}}{2}\sqrt{(l-m)(l-m-1)},\\
b^{k^\prime k}_{l,l,m+1,m}&=&
-\frac{ic_{ll}}{2}\sqrt{(l+m+1)(l-m)},\\
b^{k^\prime k}_{l+1,l,m+1,m}&=&
\frac{ic_{l+1,l}}{2}\sqrt{(l+m+1)(l+m+2)}.
\end{array}\right.}
\end{equation}
\begin{equation}\label{L3}
{\renewcommand{\arraystretch}{1.25}
\sL_3:\quad\left\{\begin{array}{ccc} c^{k^\prime k}_{l-1,l,m}&=&
c^{k^\prime k}_{l-1,l}\sqrt{l^2-m^2},\\
c^{k^\prime k}_{l,l,m}&=&c^{k^\prime k}_{ll}m,\\
c^{k^\prime k}_{l+1,l,m}&=& c^{k^\prime
k}_{l+1,l}\sqrt{(l+1)^2-m^2}.
\end{array}\right.}
\end{equation}
\begin{equation}\label{L1'}
{\renewcommand{\arraystretch}{1.25}
\sL^\ast_1:\quad\left\{\begin{array}{ccc}
d^{\dot{k}^\prime\dot{k}}_{\dot{l}-1,\dot{l},\dot{m}-1,\dot{m}}&=&
-\frac{c_{\dot{l}-1,\dot{l}}}{2}
\sqrt{(\dot{l}+\dot{m})(\dot{l}-\dot{m}-1)},\\
d^{\dot{k}^\prime\dot{k}}_{\dot{l},\dot{l},\dot{m}-1,\dot{m}}&=&
\frac{c_{\dot{l}\dot{l}}}{2}
\sqrt{(\dot{l}+\dot{m})(\dot{l}-\dot{m}+1)},\\
d^{\dot{k}^\prime\dot{k}}_{\dot{l}+1,\dot{l},\dot{m}-1,\dot{m}}&=&
\frac{c_{\dot{l}+1,\dot{l}}}{2}
\sqrt{(\dot{l}-\dot{m}+1)(\dot{l}-\dot{m}+2)},\\
d^{\dot{k}^\prime\dot{k}}_{\dot{l}-1,\dot{l},\dot{m}+1,\dot{m}}&=&
\frac{c_{\dot{l}-1,\dot{l}}}{2}
\sqrt{(\dot{l}-\dot{m})(\dot{l}-\dot{m}-1)},\\
d^{\dot{k}^\prime\dot{k}}_{\dot{l},\dot{l},\dot{m}+1,\dot{m}}&=&
\frac{c_{\dot{l}\dot{l}}}{2}
\sqrt{(\dot{l}+\dot{m}+1)(\dot{l}-\dot{m})},\\
d^{\dot{k}^\prime\dot{k}}_{\dot{l}+1,\dot{l},\dot{m}+1,\dot{m}}&=&
-\frac{c_{\dot{l}+1,\dot{l}}}{2}
\sqrt{(\dot{l}+\dot{m}+1)(\dot{l}+\dot{m}+2)}.
\end{array}\right.}
\end{equation}
\begin{equation}\label{L2'}
{\renewcommand{\arraystretch}{1.25}
\sL^\ast_2:\quad\left\{\begin{array}{ccc}
e^{\dot{k}^\prime\dot{k}}_{\dot{l}-1,\dot{l},\dot{m}-1,\dot{m}}&=&
-\frac{ic_{\dot{l}-1,\dot{l}}}{2}
\sqrt{(\dot{l}+\dot{m})(\dot{l}-\dot{m}-1)},\\
e^{\dot{k}^\prime\dot{k}}_{\dot{l},\dot{l},\dot{m}-1,\dot{m}}&=&
\frac{ic_{\dot{l}\dot{l}}}{2}
\sqrt{(\dot{l}+\dot{m})(\dot{l}-\dot{m}+1)},\\
e^{\dot{k}^\prime\dot{k}}_{\dot{l}+1,\dot{l},\dot{m}-1,\dot{m}}&=&
\frac{ic_{\dot{l}+1,\dot{l}}}{2}
\sqrt{(\dot{l}-\dot{m}+1)(\dot{l}-\dot{m}+2)},\\
e^{\dot{k}^\prime\dot{k}}_{\dot{l}-1,\dot{l},\dot{m}+1,\dot{m}}&=&
\frac{-ic_{\dot{l}-1,\dot{l}}}{2}
\sqrt{(\dot{l}-\dot{m})(\dot{l}-\dot{m}-1)},\\
e^{\dot{k}^\prime\dot{k}}_{\dot{l},\dot{l},\dot{m}+1,\dot{m}}&=&
\frac{-ic_{\dot{l}\dot{l}}}{2}
\sqrt{(\dot{l}+\dot{m}+1)(\dot{l}-\dot{m})},\\
e^{\dot{k}^\prime\dot{k}}_{\dot{l}+1,\dot{l},\dot{m}+1,\dot{m}}&=&
-\frac{ic_{\dot{l}+1,\dot{l}}}{2}
\sqrt{(\dot{l}+\dot{m}+1)(\dot{l}+\dot{m}+2)}.
\end{array}\right.}
\end{equation}
\begin{equation}\label{L3'}
{\renewcommand{\arraystretch}{1.25}
\sL^\ast_3:\quad\left\{\begin{array}{ccc}
f^{\dot{k}^\prime\dot{k}}_{\dot{l}-1,\dot{l},\dot{m}}&=&
c^{\dot{k}^\prime\dot{k}}_{\dot{l}-1,\dot{l}}
\sqrt{\dot{l}^2-\dot{m}^2},\\
f^{\dot{k}^\prime\dot{k}}_{\dot{l}\dot{l},\dot{m}}&=&
c^{\dot{k}^\prime\dot{k}}_{\dot{l}\dot{l}}\dot{m},\\
f^{\dot{k}^\prime\dot{k}}_{\dot{l}+1,\dot{l},\dot{m}}&=&
c^{\dot{k}^\prime\dot{k}}_{\dot{l}+1,\dot{l}}
\sqrt{(\dot{l}+1)^2-\dot{m}^2}.
\end{array}\right.}
\end{equation}
In general, the matrix $\sL_3$ must be a reducible representation of
the proper Lorentz group $\fG_+$, and can always be written in the
form
\begin{equation}\label{Decomp}
\sL_3=\begin{bmatrix}
\sL^{l_1}_3 & & & &\\
& \sL^{l_2}_3 & &\text{\huge 0} &\\
& & \sL^{l_3}_3 & &\\
&\text{\huge 0} & & \ddots &\\
& & & & \sL^{l_n}_3
\end{bmatrix},
\end{equation}
where $\sL^{l_i}_3$ is a spin block (the matrix
$\overset{\ast}{\sL}_3$ has the same decompositions). It is obvious
that the matrices $\sL_1$, $\sL_2$ and $\overset{\ast}{\sL}_1$,
$\overset{\ast}{\sL}_2$ admit also the decompositions of the type
(\ref{Decomp}) by definition. If the spin block $\sL^{l_i}_3$ has
non-null roots, then the particle possesses the spin $l_i$.
Corresponding to the decomposition (\ref{Decomp}), the wave function
also decomposes into a direct sum of component wave functions which
we write
\[
\boldsymbol{\psi}=\psi_{l_1m_1}+ \psi_{l_2m_2}+\psi_{l_3m_3}+\ldots.
\]
According to a de Broglie theory of fusion \cite{Bro43},
interlocking representations give rise to indecomposable
relativistic wave equations. Otherwise, we have decomposable
equations. As is known, the indecomposable wave equations correspond
to composite particles. A relation between indecomposable wave
equations and composite particles will be studied in a separate
work.

Returning to the equations (\ref{FEL}), we see that the first
equation can be written in the form
\begin{eqnarray}
\sum^3_{j=1}\Lambda^\ast_j\frac{\partial\dot{\psi}}
{\partial\widetilde{a}_j}+i\sum^3_{j=1}\Lambda^\ast_j
\frac{\partial\dot{\psi}}{\partial\widetilde{a}^\ast_j}+
\kappa^c\psi&=&0,\nonumber\\
\sum^3_{j=1}\Lambda_j\frac{\partial\psi}{\partial a_j}-
i\sum^3_{j=1}\Lambda_j\frac{\partial\psi}{\partial a^\ast_j}+
\dot{\kappa}^c\dot{\psi}&=&0,\label{Complex}
\end{eqnarray}
where $\fg_1=a_1$, $\fg_2=a_2$, $\fg_3=a_3$, $\fg_4=ia_1$,
$\fg_5=ia_2$, $\fg_6=ia_3$, $a^\ast_1=-i\fg_4$, $a^\ast_2=-i\fg_5$,
$a^\ast_3=-i\fg_6$, and $\widetilde{a}_j$, $\widetilde{a}^\ast_j$
are the parameters corresponding to the dual basis. In essence, the
equations (\ref{Complex}) are defined in a three-dimensional complex
space $\C^3$. In turn, the space $\C^3$ is isometric to a
6-dimensional bivector space $\R^6$ (a parameter space of the
Lorentz group \cite{Pet69}). The bivector space $\R^6$ is a tangent
space of the group manifold $\fL_6$ of the Lorentz group, that is,
the manifold $\fL_6$ in the each its point is equivalent locally to
the space $\R^6$. Thus, for all $\fg\in\fL_6$ we have
$T_{\fg}\fL_6\simeq\R^6$. There exists a close relationship between
the metrics of the Minkowski spacetime $\R^{1,3}$ and the metrics of
$\R^6$ (see Appendix). In the case of $\R^{1,3}$ with the metric
tensor
\[
g_{\alpha\beta}=\ar\begin{pmatrix}
-1 & 0 & 0 & 0\\
0  & -1& 0 & 0\\
0  & 0 & -1& 0\\
0  & 0 & 0 & 1
\end{pmatrix}
\]
in virtue of (\ref{Metric}) for the metric tensor of $\R^6$ we
obtain
\begin{equation}\label{MetB}
g_{ab}=\ar\begin{bmatrix}
-1& 0 & 0 & 0 & 0 & 0\\
0 & -1& 0 & 0 & 0 & 0\\
0 & 0 & -1& 0 & 0 & 0\\
0 & 0 & 0 & 1 & 0 & 0\\
0 & 0 & 0 & 0 & 1 & 0\\
0 & 0 & 0 & 0 & 0 & 1
\end{bmatrix},
\end{equation}
where the order of collective indices in $\R^6$ is $23\rightarrow
0$, $10\rightarrow 1$, $20\rightarrow 2$, $30\rightarrow 3$,
$31\rightarrow 4$, $12\rightarrow 5$. The system (\ref{Complex}) in
the bivector space is written as follows:
\begin{multline}
\sum_i\left[
g^-_{i1}\sL_1T^{-1}_{\fg}\frac{\partial\psi^\prime}{\partial
a^\prime_i}+
g^-_{i2}\sL_2T^{-1}_{\fg}\frac{\partial\psi^\prime}{\partial
a^\prime_i}+
g^-_{i3}\sL_3T^{-1}_{\fg}\frac{\partial\psi^\prime}{\partial
a^\prime_i}-
\right.\\
\shoveright{\left.
-ig^-_{i1}\sL_1T^{-1}_{\fg}\frac{\partial\psi^\prime} {\partial
a^\ast_i{}^\prime}-
ig^-_{i2}\sL_2T^{-1}_{\fg}\frac{\partial\psi^\prime} {\partial
a^\ast_i{}^\prime}-
ig^-_{i3}\sL_3T^{-1}_{\fg}\frac{\partial\psi^\prime} {\partial
a^\ast_i{}^\prime}\right]+
\kappa^cT^{-1}_{\fg}\psi^\prime=0,}\\
\shoveleft{ \sum_i\left[
g^+_{i1}\sL^\ast_1\overset{\ast}{T}_{\fg}\!\!{}^{-1}
\frac{\partial\dot{\psi}^\prime}{\partial\widetilde{a}^\prime_i}+
g^+_{i2}\sL^\ast_2\overset{\ast}{T}_{\fg}\!\!{}^{-1}
\frac{\partial\dot{\psi}^\prime}{\partial\widetilde{a}^\prime_i}+
g^+_{i3}\sL^\ast_3\overset{\ast}{T}_{\fg}\!\!{}^{-1}
\frac{\partial\dot{\psi}^\prime}{\partial\widetilde{a}^\prime_i}+\right.}\\
\left. +ig^+_{i1}\sL^\ast_1\overset{\ast}{T}_{\fg}\!\!{}^{-1}
\frac{\partial\dot{\psi}^\prime}{\partial\widetilde{a^\ast}_i{}^\prime}+
ig^+_{i2}\sL^\ast_2\overset{\ast}{T}_{\fg}\!\!{}^{-1}
\frac{\partial\dot{\psi}^\prime}{\partial\widetilde{a^\ast}_i{}^\prime}+
ig^+_{i3}\sL^\ast_3\overset{\ast}{T}_{\fg}\!\!{}^{-1}
\frac{\partial\dot{\psi}^\prime}{\partial\widetilde{a^\ast}_i{}^\prime}\right]+
\dot{\kappa}^c\overset{\ast}{T}_{\fg}\!\!{}^{-1}\dot{\psi}^\prime=0,\nonumber
\end{multline}
where
\begin{eqnarray}
\frac{\partial}{\partial a_1}&=&-\frac{\sin\varphi^c}{r\sin\theta^c}
\frac{\partial}{\partial\varphi}+\frac{\cos\varphi^c\cos\theta^c}{r}
\frac{\partial}{\partial\theta}+\cos\varphi^c\sin\theta^c\frac{\partial}
{\partial r},\label{CD1}\\
\frac{\partial}{\partial a_2}&=&\frac{\cos\varphi^c}{r\sin\theta^c}
\frac{\partial}{\partial\varphi}+\frac{\sin\varphi^c\cos\theta^c}{r}
\frac{\partial}{\partial\theta}+\sin\varphi^c\sin\theta^c\frac{\partial}
{\partial r},\label{CD2}\\
\frac{\partial}{\partial
a_3}&=&-\frac{\sin\theta^c}{r}\frac{\partial}
{\partial\theta}+\cos\theta^c\frac{\partial}{\partial r}.\label{CD3}
\end{eqnarray}
\begin{eqnarray}
\frac{\partial}{\partial a^\ast_1}&=&i\frac{\partial}{\partial a_1}=
-\frac{\sin\varphi^c}{r\sin\theta^c}\frac{\partial}{\partial\epsilon}+
\frac{\cos\varphi^c\sin\theta^c}{r}\frac{\partial}{\partial\tau}+
i\cos\varphi^c\sin\theta^c\frac{\partial}{\partial r},\label{CD4}\\
\frac{\partial}{\partial a^\ast_2}&=&i\frac{\partial}{\partial a_2}=
\frac{\cos\varphi^c}{r\sin\theta^c}\frac{\partial}{\partial\epsilon}+
\frac{\sin\varphi^c\cos\theta^c}{r}\frac{\partial}{\partial\tau}+
i\sin\varphi^c\sin\theta^c\frac{\partial}{\partial r},\label{CD5}\\
\frac{\partial}{\partial a^\ast_3}&=&i\frac{\partial}{\partial a_3}=
-\frac{\sin\theta^c}{r}\frac{\partial}{\partial\tau}+
i\cos\theta^c\frac{\partial}{\partial r}.\label{CD6}
\end{eqnarray}
\begin{eqnarray}
\frac{\partial}{\partial\widetilde{a}_1}&=&
-\frac{\sin\dot{\varphi}^c}{r^\ast\sin\dot{\theta}^c}\frac{\partial}
{\partial\varphi}+\frac{\cos\dot{\varphi}^c\cos\dot{\theta}^c}{r^\ast}
\frac{\partial}{\partial\theta}+\cos\dot{\varphi}^c\sin\dot{\theta}^c
\frac{\partial}{\partial r^\ast},\label{CDD1}\\
\frac{\partial}{\partial\widetilde{a}_2}&=&
\frac{\cos\dot{\varphi}^c}{r^\ast\sin\dot{\theta}^c}\frac{\partial}
{\partial\varphi}+\frac{\sin\dot{\varphi}^c\cos\dot{\theta}^c}{r^\ast}
\frac{\partial}{\partial\theta}+\sin\dot{\varphi}^c\sin\dot{\theta}^c
\frac{\partial}{\partial r^\ast},\label{CDD2}\\
\frac{\partial}{\partial\widetilde{a}_3}&=&
-\frac{\sin\dot{\theta}^c}{r^\ast}\frac{\partial}{\partial\theta}+
\cos\dot{\theta}^c\frac{\partial}{\partial r^\ast}.\label{CDD3}
\end{eqnarray}
\begin{eqnarray}
\frac{\partial}{\partial\widetilde{a}^\ast_1}&=&
-i\frac{\partial}{\partial\widetilde{a}_1}=
\frac{\sin\dot{\varphi}^c}{r^\ast\sin\dot{\theta}^c}\frac{\partial}
{\partial\epsilon}-\frac{\cos\dot{\varphi}^c\cos\dot{\theta}^c}{r^\ast}
\frac{\partial}{\partial\tau}-i\cos\dot{\varphi}^c\sin\dot{\theta}^c
\frac{\partial}{\partial r^\ast},\label{CDD4}\\
\frac{\partial}{\partial\widetilde{a}^\ast_2}&=&
-i\frac{\partial}{\partial\widetilde{a}_2}=
-\frac{\cos\dot{\varphi}^c}{r^\ast\sin\dot{\theta}^c}\frac{\partial}
{\partial\epsilon}-\frac{\sin\dot{\varphi}^c\cos\dot{\theta}^c}{r^\ast}
\frac{\partial}{\partial\tau}-i\sin\dot{\varphi}^c\sin\dot{\theta}^c
\frac{\partial}{\partial r^\ast},\label{CDD5}\\
\frac{\partial}{\partial\widetilde{a}^\ast_3}&=&
-i\frac{\partial}{\partial\widetilde{a}_3}=
\frac{\sin\dot{\theta}^c}{r^\ast}\frac{\partial}{\partial\tau}
-i\cos\dot{\theta}^c\frac{\partial}{\partial r^\ast}\label{CDD6}
\end{eqnarray}
are derivatives defined on the spheres (\ref{CS}) and (\ref{DS}).
Solutions of wave equations for extended object are found in the
form of series in associated hyperspherical functions which defined
on the spheres $S^2_c$ and $\dot{S}^2_c$:
\begin{eqnarray}
\psi^k_{lm;\dot{l}\dot{m}}&=&\boldsymbol{f}^{l_0}_{lmk}(r)
\fM^{l_0}_{mn}(\varphi,\epsilon,\theta,\tau,0,0),\nonumber\\
\psi^{\dot{k}}_{\dot{l}\dot{m};lm}&=&
\boldsymbol{f}^{\dot{l}_0}_{\dot{l}\dot{m}\dot{k}}(r^\ast)
\fM^{\dot{l}_0}_{\dot{m}\dot{n}}(\varphi,\epsilon,\theta,\tau,0,0),\nonumber
\end{eqnarray}
where $l_0\ge l$, $-l_0\le m$, $n\le l_0$ and $\dot{l}_0\ge\dot{l}$,
$-\dot{l}_0\le\dot{m}$, $\dot{n}\le\dot{l}_0$.

\emph{We claim that a system of Dirac like wave equations}
\begin{equation}\label{WE}
{\renewcommand{\arraystretch}{1.45} \left\{\begin{array}{cll}
\Gamma_i\dfrac{\partial\psi(x)}{\partial x_i}+\kappa\psi(x)&=&0,\\
\Gamma^T_i\dfrac{\partial\overline{\psi}(x)}{\partial x_i}-
\kappa\overline{\psi}(x)&=&0, \quad(i=0,\ldots,3);\\
\Gamma_k\dfrac{\partial\psi(\fg)}{\partial\fg_k}+\kappa\psi(\fg)&=&0,\\
\Gamma^T_k\dfrac{\overline{\psi}(\fg)}{\partial\fg_k}-
\kappa\overline{\psi}(\fg)&=&0 \quad(k=1,\ldots,6)
\end{array}
\right.}
\end{equation}
\emph{describes extended particles of any spin on the group manifold
$\cM_{10}=\R^{1,3}\times\fL_6$ of the Poincar\'{e} group $\cP$} (in
particular, on the homogeneous space $\cM_8=\R^{1,3}\times S^2_c$).
In its turn, the Ginzburg-Tamm system of Klein-Gordon like wave
equations (\ref{GT2}) describes extended particles on the
homogeneous space $\cM=\R^{1,3}\times H^3$ of $\cP$, where $H^3$ is
a 3-dimensional one-sheeted hyperboloid. A relation between the
systems (\ref{WE}) and (\ref{GT2}) is similar to a relation between
usual Dirac and Klein-Gordon wave equations for the point-like
particles. Namely, equations (\ref{WE}) (differential equations of
the first order) can be understood as `square roots' of the
equations (\ref{GT2}) (differential equations of the second order).

\subsection{The Dirac field}
In accordance with the general Fermi-scheme (Fig.\,1) of the
interlocking representations of $\fG_+$ the field
$(1/2,0)\oplus(0,1/2)$ is defined within the following chain:
\[
\dgARROWLENGTH=2.5em
\begin{diagram}
\node{\left(\frac{1}{2},0\right)}\arrow{e,<>}
\node{\left(0,\frac{1}{2}\right)}
\end{diagram}.
\]
We start with the Lagrangian (\ref{Lagrange}) on the group manifold
$\cM_{10}$:
\begin{multline}\label{LagDir}
\cL(\balpha)=-\frac{1}{2}\left(\overline{\boldsymbol{\psi}}(\balpha)\Gamma_\mu
\frac{\partial\boldsymbol{\psi}(\balpha)}{\partial x_\mu}-
\frac{\partial\overline{\boldsymbol{\psi}}(\balpha)}{\partial x_\mu}
\Gamma_\mu\boldsymbol{\psi}(\balpha)\right)-\\
-\frac{1}{2}\left(\overline{\boldsymbol{\psi}}(\balpha)\Upsilon_\nu
\frac{\partial\boldsymbol{\psi}(\balpha)}{\partial\fg_\nu}-
\frac{\partial\overline{\boldsymbol{\psi}}(\balpha)}{\partial\fg_\nu}
\Upsilon_\nu\boldsymbol{\psi}(\balpha)\right)-
\kappa\overline{\boldsymbol{\psi}}(\balpha)\boldsymbol{\psi}(\balpha),
\end{multline}
where $\boldsymbol{\psi}(\balpha)=\psi(x)\psi(\fg)$
($\mu=0,1,2,3,\;\nu=1,\ldots,6$), and
\begin{equation}\label{Gamma1}
\gamma_0=\begin{pmatrix}
\sigma_0 & 0\\
0 & -\sigma_0
\end{pmatrix},\;\;\gamma_1=\begin{pmatrix}
0 & \sigma_1\\
-\sigma_1 & 0
\end{pmatrix},\;\;\gamma_2=\begin{pmatrix}
0 & \sigma_2\\
-\sigma_2 & 0
\end{pmatrix},\;\;\gamma_3=\begin{pmatrix}
0 & \sigma_3\\
-\sigma_3 & 0
\end{pmatrix},
\end{equation}
\begin{equation}\label{Upsilon1}
\Upsilon_1=\begin{pmatrix}
0 & \Lambda^\ast_1\\
\Lambda_1 & 0
\end{pmatrix},\quad\Upsilon_2=\begin{pmatrix}
0 & \Lambda^\ast_2\\
\Lambda_2 & 0
\end{pmatrix},\quad\Upsilon_3=\begin{pmatrix}
0 & \Lambda^\ast_3\\
\Lambda_3 & 0
\end{pmatrix},
\end{equation}
\begin{equation}\label{Upsilon2}
\Upsilon_4=\begin{pmatrix}
0 & i\Lambda^\ast_1\\
i\Lambda_1 & 0
\end{pmatrix},\quad\Upsilon_5=\begin{pmatrix}
0 & i\Lambda^\ast_2\\
i\Lambda_2 & 0
\end{pmatrix},\quad\Upsilon_6=\begin{pmatrix}
0 & i\Lambda^\ast_3\\
i\Lambda_3 & 0
\end{pmatrix},
\end{equation}
where $\sigma_i$ are the Pauli matrices, and the matrices
$\Lambda_j$ and $\Lambda^\ast_j$ are derived from
(\ref{L1})--(\ref{L3}) and (\ref{L1'})--(\ref{L3'}) at $l=1/2$:
\begin{eqnarray}
&&\sL_1=\frac{1}{2}c_{\frac{1}{2}\frac{1}{2}}\begin{pmatrix}
0 & 1\\
1 & 0
\end{pmatrix},\quad
\sL_2=\frac{1}{2}c_{\frac{1}{2}\frac{1}{2}}\begin{pmatrix}
0 & -i\\
i & 0
\end{pmatrix},\quad
\sL_3=\frac{1}{2}c_{\frac{1}{2}\frac{1}{2}}\begin{pmatrix}
1 & 0\\
0 & -1
\end{pmatrix},\nonumber\\
&&\sL^\ast_1=\frac{1}{2}\dot{c}_{\frac{1}{2}\frac{1}{2}}\begin{pmatrix}
0 & 1\\
1 & 0
\end{pmatrix},\quad
\sL^\ast_2=\frac{1}{2}\dot{c}_{\frac{1}{2}\frac{1}{2}}\begin{pmatrix}
0 & -i\\
i & 0
\end{pmatrix},\quad
\sL^\ast_3=\frac{1}{2}\dot{c}_{\frac{1}{2}\frac{1}{2}}\begin{pmatrix}
1 & 0\\
0 & -1
\end{pmatrix}.\label{LDir}
\end{eqnarray}
It is easy to see that these matrices coincide with the Pauli
matrices $\sigma_i$ when $c_{\frac{1}{2}\frac{1}{2}}=2$. The Dirac
bispinor $\psi=(\psi_1,\psi_2,\dot{\psi}_1,\dot{\psi}_2)^T$ is
defined on $\cM_8=\R^{1,3}\times S^2_c$ by the following components
\cite{Var04e}:
\begin{eqnarray}
&&\psi^l_{1n}(\balpha)=\psi^+_1(x)\psi^l_{1n}(\fg)= u_1(\bp)e^{-i
px}\boldsymbol{f}^l_{\frac{1}{2},\frac{1}{2}}
(\re r)\fM^l_{\frac{1}{2},n}(\varphi,\epsilon,\theta,\tau,0,0),\nonumber\\
&&\psi^l_{2n}(\balpha)=\psi^+_2(x)\psi^l_{2n}(\fg)= \pm
u_2(\bp)e^{-ipx}\boldsymbol{f}^l_{\frac{1}{2},\frac{1}{2}}
(\re r)\fM^l_{-\frac{1}{2},n}(\varphi,\epsilon,\theta,\tau,0,0),\nonumber\\
&&\dot{\psi}^{\dot{l}}_{1\dot{n}}(\balpha)=
\psi^-_1(x)\dot{\psi}^{\dot{l}}_{1\dot{n}}(\fg)= \mp
v_1(\bp)e^{ipx}\boldsymbol{f}^{\dot{l}}_{\frac{1}{2},-\frac{1}{2}}
(\re r^\ast)\fM^{\dot{l}}_{\frac{1}{2},\dot{n}}
(\varphi,\epsilon,\theta,\tau,0,0),\nonumber\\
&&\dot{\psi}^{\dot{l}}_{2\dot{n}}(\balpha)=
\psi^-_2(x)\dot{\psi}^{\dot{l}}_{2\dot{n}}(\fg)=
v_2(\bp)e^{ipx}\boldsymbol{f}^{\dot{l}}_{\frac{1}{2},-\frac{1}{2}}
(\re r^\ast)\fM^{\dot{l}}_{-\frac{1}{2},\dot{n}}
(\varphi,\epsilon,\theta,\tau,0,0), \nonumber
\end{eqnarray}
where
\begin{gather}
u_1(\bp)=\left(\frac{E+m}{2m}\right)^{1/2}\ar\begin{pmatrix}
1\\
0\\
\frac{p_z}{E+m}\\
\frac{p_+}{E+m}
\end{pmatrix},\quad
u_2(\bp)=\left(\frac{E+m}{2m}\right)^{1/2}\begin{pmatrix}
0\\
1\\
\frac{p_-}{E+m}\\
\frac{-p_z}{E+m}
\end{pmatrix},\nonumber\\
v_1(\bp)=\left(\frac{E+m}{2m}\right)^{1/2}\ar\begin{pmatrix}
\frac{p_z}{E+m}\\
\frac{p_+}{E+m}\\
1\\
0
\end{pmatrix},\quad
v_2(\bp)=\left(\frac{E+m}{2m}\right)^{1/2}\begin{pmatrix}
\frac{p_-}{E+m}\\
\frac{-p_z}{E+m}\\
0\\
1
\end{pmatrix},\nonumber
\end{gather}
here $p_\pm=p_x\pm ip_y$. Radial functions have the form
\[
\boldsymbol{f}^l_{\frac{1}{2},\frac{1}{2}}(\re r)=
C_1\sqrt{\kappa^c\dot{\kappa}^c}\re r
J_l\left(\sqrt{\kappa^c\dot{\kappa}^c}\re r\right)+
C_2\sqrt{\kappa^c\dot{\kappa}^c}\re r
J_{-l}\left(\sqrt{\kappa^c\dot{\kappa}^c}\re r\right),
\]
\[
\boldsymbol{f}^{\dot{l}}_{\frac{1}{2},-\frac{1}{2}}(\re r^\ast)=
\frac{C_1}{2}\sqrt{\frac{\dot{\kappa}^c}{\kappa^c}}\re r^\ast
J_{l+1}\left(\sqrt{\kappa^c\dot{\kappa}^c}\re r^\ast\right)
-\frac{C_2}{2}\sqrt{\frac{\dot{\kappa}^c}{\kappa^c}}\re r^\ast
J_{-l-1}\left(\sqrt{\kappa^c\dot{\kappa}^c}\re r^\ast\right),
\]
where $J_l\left(\sqrt{\kappa^c\dot{\kappa}^c}\re r\right)$ are the
Bessel functions of half-integer order, and
\begin{eqnarray}
&&l=\frac{1}{2},\;\frac{3}{2},\;\frac{5}{2},\ldots;\nonumber\\
&&\dot{l}=\frac{1}{2},\;\frac{3}{2},\;\frac{5}{2},\ldots;\nonumber
\end{eqnarray}
\[
\fM_l^{\pm\frac{1}{2}}(\varphi,\epsilon,\theta,\tau,0,0)=
e^{\mp\frac{1}{2}(\epsilon+i\varphi)}\fZ_l^{\pm\frac{1}{2}}(\theta,\tau),
\]
\begin{multline}
\fZ_l^{\pm\frac{1}{2}}(\theta,\tau)=\cos^{2l}\frac{\theta}{2}
\cosh^{2l}\frac{\tau}{2}\sum^l_{k=-l}i^{\pm\frac{1}{2}-k}
\tan^{\pm\frac{1}{2}-k}\frac{\theta}{2}\tanh^{-k}\frac{\tau}{2}\times\\
\hypergeom{2}{1}{\pm\frac{1}{2}-l+1,1-l-k}{\pm\frac{1}{2}-k+1}
{i^2\tan^2\frac{\theta}{2}}
\hypergeom{2}{1}{-l+1,1-l-k}{-k+1}{\tanh^2\frac{\tau}{2}},\nonumber
\end{multline}
\[
\fM_{\dot{l}}^{\pm\frac{1}{2}}(\varphi,\epsilon,\theta,\tau,0,0)=
e^{\mp\frac{1}{2}(\epsilon-i\varphi)}
\fZ_{\dot{l}}^{\pm\frac{1}{2}}(\theta,\tau),
\]
\begin{multline}
\fZ_{\dot{l}}^{\pm\frac{1}{2}}(\theta,\tau)=
\cos^{2\dot{l}}\frac{\theta}{2} \cosh^{2\dot{l}}\frac{\tau}{2}
\sum^{\dot{l}}_{\dot{k}=-\dot{l}}i^{\pm\frac{1}{2}-\dot{k}}
\tan^{\pm\frac{1}{2}-\dot{k}}\frac{\theta}{2}
\tanh^{-\dot{k}}\frac{\tau}{2}\times\\
\hypergeom{2}{1}{\pm\frac{1}{2}-\dot{l}+1,1-\dot{l}-\dot{k}}
{\pm\frac{1}{2}-\dot{k}+1} {i^2\tan^2\frac{\theta}{2}}
\hypergeom{2}{1}{-\dot{l}+1,1-\dot{l}-\dot{k}}
{-\dot{k}+1}{\tanh^2\frac{\tau}{2}}\nonumber
\end{multline}
are the associated hyperspherical functions defined on the spheres
$S^c_2$ and $\dot{S}^c_2$. General solutions are found in the form
of generalized Fourier integrals
\begin{eqnarray}
\psi_1(\balpha)&=&\sum^\infty_{l=\frac{1}{2}}\boldsymbol{f}^l_{\frac{1}{2},\frac{1}{2}}(\re
r) \sum^l_{n=-l}\int\limits_{T_4}u_1(\bp)e^{ipx}
\alpha^{\frac{1}{2}}_{l,n}
\fM^l_{\frac{1}{2},n}(\varphi,\epsilon,\theta,\tau,0,0)d^4x,\nonumber\\
\psi_2(\balpha)&=&\pm\sum^\infty_{l=\frac{1}{2}}\boldsymbol{f}^l_{\frac{1}{2},\frac{1}{2}}(\re
r) \sum^l_{n=-l}\int\limits_{T_4}u_2(\bp)e^{ipx}
\alpha^{-\frac{1}{2}}_{l,n}
\fM^l_{-\frac{1}{2},n}(\varphi,\epsilon,\theta,\tau,0,0)d^4x,\nonumber\\
\dot{\psi}_1(\balpha)&=&\mp \sum^\infty_{\dot{l}=\frac{1}{2}}
\boldsymbol{f}^{\dot{l}}_{\frac{1}{2},-\frac{1}{2}}(\re r^\ast)
\sum^{\dot{l}}_{\dot{n}=-\dot{l}}\int\limits_{T_4}v_1(\bp)e^{-ipx}
\alpha^{\frac{1}{2}}_{\dot{l},\dot{n}}
\fM^{\dot{l}}_{\frac{1}{2},\dot{n}}
(\varphi,\epsilon,\theta,\tau,0,0)d^4x,\nonumber\\
\dot{\psi}_2(\balpha)&=&\sum^\infty_{\dot{l}=\frac{1}{2}}
\boldsymbol{f}^{\dot{l}}_{\frac{1}{2},-\frac{1}{2}}(\re r^\ast)
\sum^{\dot{l}}_{\dot{n}=-\dot{l}}\int\limits_{T_4}v_2(\bp)e^{-ipx}
\alpha^{-\frac{1}{2}}_{\dot{l},\dot{n}}
\fM^{\dot{l}}_{-\frac{1}{2},\dot{n}}
(\varphi,\epsilon,\theta,\tau,0,0)d^4x,\nonumber
\end{eqnarray}
where
\begin{eqnarray}
\alpha^{\pm\frac{1}{2}}_{l,n}&=&
\frac{(-1)^n(2l+1)(2\dot{l}+1)}{32\pi^4
\boldsymbol{f}^l_{\frac{1}{2},\frac{1}{2}}(\re a)}
\int\limits_{S^2_c}\int\limits_{T_4}
F_{\pm\frac{1}{2}}(\balpha)e^{-ipx}
\fM^l_{\pm\frac{1}{2},n}(\varphi,\epsilon,\theta,\tau,0,0)d^4xd^4\fg,\nonumber\\
\alpha^{\pm\frac{1}{2}}_{\dot{l},\dot{n}}&=&
\frac{(-1)^{\dot{n}}(2l+1)(2\dot{l}+1)}{32\pi^4
\boldsymbol{f}^{\dot{l}}_{\frac{1}{2},-\frac{1}{2}}(\re a^\ast)}
\int\limits_{S^2_c}\int\limits_{T_4}
\dot{F}_{\pm\frac{1}{2}}(\balpha)e^{ipx}
\fM^{\dot{l}}_{\pm\frac{1}{2},\dot{n}}
(\varphi,\epsilon,\theta,\tau,0,0)d^4xd^4\fg.\nonumber
\end{eqnarray}
\subsection{The Maxwell field} In accordance with the general
Bose-scheme of the interlocking representations of $\fG_+$ (see
Fig.\,1), the field $(1,0)\oplus(0,1)$ is defined within the
following interlocking scheme:
\[
\dgARROWLENGTH=2.5em
\begin{diagram}
\node{(1,0)}\arrow{e,<>}\node{\left(\frac{1}{2},\frac{1}{2}\right)}
\arrow{e,<>}\node{(0,1)}
\end{diagram}.
\]
We start with the Lagrangian (\ref{Lagrange}) on the group manifold
$\cM_{10}$. Let us rewrite (\ref{Lagrange}) in the form
\begin{multline}\label{LagMax}
\cL(\balpha)=-\frac{1}{2}\left(\overline{\boldsymbol{\phi}}(\balpha)\Gamma_\mu
\frac{\partial\boldsymbol{\phi}(\balpha)}{\partial x_\mu}-
\frac{\partial\overline{\boldsymbol{\phi}}(\balpha)}{\partial x_\mu}
\Gamma_\mu\boldsymbol{\phi}(\balpha)\right)-\\
-\frac{1}{2}\left(\overline{\boldsymbol{\phi}}(\balpha)\Upsilon_\nu
\frac{\partial\boldsymbol{\phi}(\balpha)}{\partial\fg_\nu}-
\frac{\partial\overline{\boldsymbol{\phi}}(\balpha)}{\partial\fg_\nu}
\Upsilon_\nu\boldsymbol{\phi}(\balpha)\right),
\end{multline}
where $\boldsymbol{\phi}(\balpha)=\phi(x)\phi(\fg)$
($\mu=0,1,2,3,\;\nu=1,\ldots,6$), and
\begin{equation}\label{Gamma2}
\Gamma_0=\begin{pmatrix}
0 & I\\
I & 0
\end{pmatrix},\;\;\Gamma_1=\begin{pmatrix}
0 & -\alpha_1\\
\alpha_1 & 0
\end{pmatrix},\;\;\Gamma_2=\begin{pmatrix}
0 & -\alpha_2\\
\alpha_2 & 0
\end{pmatrix},\;\;\Gamma_3=\begin{pmatrix}
0 & -\alpha_3\\
\alpha_3 & 0
\end{pmatrix},
\end{equation}
\begin{equation}\label{Upsilon3}
\Upsilon_1=\begin{pmatrix}
0 & \Lambda^\ast_1\\
\Lambda_1 & 0
\end{pmatrix},\quad\Upsilon_2=\begin{pmatrix}
0 & \Lambda^\ast_2\\
\Lambda_2 & 0
\end{pmatrix},\quad\Upsilon_3=\begin{pmatrix}
0 & \Lambda^\ast_3\\
\Lambda_3 & 0
\end{pmatrix},
\end{equation}
\begin{equation}\label{Upsilon4}
\Upsilon_4=\begin{pmatrix}
0 & i\Lambda^\ast_1\\
i\Lambda_1 & 0
\end{pmatrix},\quad\Upsilon_5=\begin{pmatrix}
0 & i\Lambda^\ast_2\\
i\Lambda_2 & 0
\end{pmatrix},\quad\Upsilon_6=\begin{pmatrix}
0 & i\Lambda^\ast_3\\
i\Lambda_3 & 0
\end{pmatrix},
\end{equation}
where
\begin{equation}\label{Alpha}
\alpha_1=\begin{pmatrix}
0 & 0 & 0\\
0 & 0 & i\\
0 & -i& 0
\end{pmatrix},\quad\alpha_2=\begin{pmatrix}
0 & 0 & -i\\
0 & 0 & 0\\
i & 0 & 0
\end{pmatrix},\quad\alpha_3=\begin{pmatrix}
0 & i & 0\\
-i& 0 & 0\\
0 & 0 & 0
\end{pmatrix},
\end{equation}
and the matrices $\Lambda_j$ and $\Lambda^\ast_j$ are derived from
(\ref{L1})--(\ref{L3}) and (\ref{L1'})--(\ref{L3'}) at $l=1$:
\begin{equation}\label{Lambda1}
\Lambda_1=\frac{c_{11}}{\sqrt{2}}\begin{pmatrix}
0 & 1 & 0\\
1 & 0 & 0\\
0 & 1 & 0
\end{pmatrix},\quad\Lambda_2=\frac{c_{11}}{\sqrt{2}}\begin{pmatrix}
0 & -i & 0\\
i & 0 & -i\\
0 & i & 0
\end{pmatrix},\quad\Lambda_3=c_{11}\begin{pmatrix}
1 & 0 & 0\\
0 & 0 & 0\\
0 & 0 &-1
\end{pmatrix}.
\end{equation}
\begin{equation}\label{Lambda2}
\sL^\ast_1=\frac{\sqrt{2}}{2}\dot{c}_{11}\begin{pmatrix}
0 & 1 & 0\\
1 & 0 & 1\\
0 & 1 & 0
\end{pmatrix},\quad
\sL^\ast_2=\frac{\sqrt{2}}{2}\dot{c}_{11}\begin{pmatrix}
0 & -i & 0\\
i & 0  & -i\\
0 & i & 0
\end{pmatrix},\quad
\sL^\ast_3=\dot{c}_{11}\begin{pmatrix}
1 & 0 & 0\\
0 & 0 & 0\\
0 & 0 & -1
\end{pmatrix}.
\end{equation}
At this point, electromagnetic field should be defined in the
Riemann-Silberstein representation \cite{Web01,Sil07,Bir96}. The
Riemann-Silberstein (Majorana-Oppenheimer) representation considered
during long time by many authors
\cite{Maj,Opp31,Goo57,Mos59,SS62,MRB74,DaS79,Gia85}. The interest to
this formulation of electrodynamics has grown in recent years
\cite{Ina94,Sip95,Ger98,Esp98}. One of the main advantages of this
approach lies in the fact that Dirac and Maxwell fields are derived
similarly from the Dirac-like Lagrangians. These fields have the
analogous mathematical structure, namely, they are the functions on
the Poincar\'{e} group. This circumstance allows us to consider the
fields $(1/2,0)\oplus(0,1/2)$ and $(1,0)\oplus(0,1)$ on an equal
footing, from the one group theoretical viewpoint.

In 1949, Newton and Wigner \cite{NW49} showed that for the photon
there exist no localized states. Therefore, a massless field
$(1,0)\oplus(0,1)$ should be considered within the unitary
infinite-dimensional representation of the Lorentz group. In
accordance with the Naimark theorem \cite{Nai58}, the representation
$S_k$ of the subgroup $\SU(2)$ is contained in $\fS_{\lambda,\rho}$
not more then one time. Matrix elements of the representation
$\fS_{\lambda,\rho}$ are defined by the functions (\ref{MPrincip}).
\emph{We suppose that the photon field is described within the
infinite-dimensional representation $\fS_{\lambda,\rho}$}. This
representation contains a finite-dimensional representation
associated with the field $(1,0)\oplus(0,1)$.

Since longitudinal solutions
$\phi^{-\frac{1}{2}+i\rho,l_0}_{0,n}(\balpha)$ and
$\dot{\phi}^{-\frac{1}{2}-i\rho,l_0}_{0,\dot{n}}(\balpha)$ do not
contribute to a real photon due to their transversality conditions,
then particular solutions for the Majorana-Oppenheimer bispinor
$\boldsymbol{\phi}=(\phi_1,\phi_2,\phi_3,\dot{\phi}_1,\dot{\phi}_2,
\dot{\phi}_3)^T$ on $\cM_8$ are defined by the following expressions
\cite{Var05b}:
\begin{multline}
\phi^{-\frac{1}{2}+i\rho,l_0}_{\pm 1,n}(\balpha)=\phi_{\pm}(\bk;\bx,t)\phi^{-\frac{1}{2}+i\rho,l_0}_{\pm 1,n}(\fg)=\\
\lf 2(2\pi)^3\rf^{-\frac{1}{2}}
\ar\begin{pmatrix}\varepsilon_\pm(\bk)\\
\varepsilon_\pm(\bk)\end{pmatrix} \exp[i(\bk\cdot\bx-\omega
t)]\boldsymbol{f}^{-\frac{1}{2}+i\rho,l_0}_{1,\pm 1}(r) \fM_{\pm
1,n}^{-\frac{1}{2}+i\rho,l_0}(\varphi,\epsilon,\theta,\tau,0,0),\nonumber
\end{multline}
\begin{multline}
\dot{\phi}^{-\frac{1}{2}-i\rho,\dot{l}_0}_{\pm 1,\dot{n}}(\balpha)=
\phi^\ast_\pm(\bk;\bx,t)\dot{\phi}^{-\frac{1}{2}-i\rho,\dot{l}_0}_{\pm 1,\dot{n}}(\fg)=\\
\lf 2(2\pi)^3\rf^{-\frac{1}{2}}
\ar\begin{pmatrix}\varepsilon^\ast_\pm(\bk)\\
\varepsilon^\ast_\pm(\bk)\end{pmatrix} \exp[-i(\bk\cdot\bx-\omega
t)]\boldsymbol{f}^{-\frac{1}{2}-i\rho,\dot{l}_0}_{1,\pm 1}(r^\ast)
\fM_{\pm
1,\dot{n}}^{-\frac{1}{2}-i\rho,\dot{l}_0}(\varphi,\epsilon,\theta,\tau,0,0),\nonumber
\end{multline}
where
\begin{eqnarray}
\varepsilon_\pm(\bk)&=&\lf 2|\bk|^2(k^2_1+k^2_2)\rf^{-\frac{1}{2}}
\begin{bmatrix}
-k_1k_3\pm ik_2|\bk|\\
-k_2k_3\mp ik_1|\bk|\\
k^2_1+k^2_2
\end{bmatrix}\nonumber
\end{eqnarray}
are the polarization vectors of a photon. Radial functions have the
form
\begin{eqnarray}
\boldsymbol{f}^{-\frac{1}{2}+i\rho,l_0}_{1,1}(r)&=&C\sqrt{r}+
\sqrt{\left(l_0+i\rho+\frac{1}{2}\right)\left(\frac{l_0}{2}+i\frac{\rho}{2}+\frac{5}{4}\right)}r,\nonumber\\
\boldsymbol{f}^{-\frac{1}{2}-i\rho,\dot{l}_0}_{1,1}(r^\ast)&=&\dot{C}\sqrt{r^\ast}+
\sqrt{\left(\dot{l}_0-i\rho+\frac{3}{2}\right)\left(\frac{\dot{l}_0}{2}-i\frac{\rho}{2}+\frac{7}{4}\right)}r^\ast,\nonumber
\end{eqnarray}
and
\[
\fM^{-\frac{1}{2}+i\rho,l_0}_{\pm
1,n}(\varphi,\epsilon,\theta,\tau,0,0)=
e^{\mp(\epsilon+i\varphi)}\fZ^{-\frac{1}{2}+i\rho,l_0}_{\pm
1,n}(\theta,\tau),
\]
\begin{multline}
\fZ^{-\frac{1}{2}+i\rho,l_0}_{\pm
1,n}(\theta,\tau)=\cos^{2l_0}\frac{\theta}{2}
\cosh^{-1+2i\rho,l_0}\frac{\tau}{2}\sum^{l_0}_{k=-l_0}i^{\pm 1-k}
\tan^{\pm 1-k}\frac{\theta}{2}\tanh^{n-k}\frac{\tau}{2}\times\\
\hypergeom{2}{1}{\pm 1-l_0+1,1-l_0-k}{\pm 1-k+1}
{i^2\tan^2\frac{\theta}{2}}
\hypergeom{2}{1}{n-i\rho+\frac{1}{2},\frac{1}{2}-i\rho-k}{n-k+1}{\tanh^2\frac{\tau}{2}},\nonumber
\end{multline}
\[
\fM^{-\frac{1}{2}-i\rho,l_0}_{\pm
1,\dot{n}}(\varphi,\epsilon,\theta,\tau,0,0)=
e^{\mp(\epsilon-i\varphi)} \fZ^{-\frac{1}{2}-i\rho,l_0}_{\pm
1,\dot{n}}(\theta,\tau),
\]
\begin{multline}
\fZ^{-\frac{1}{2}-i\rho,l_0}_{\pm 1,\dot{n}}(\theta,\tau)=
\cos^{2l_0}\frac{\theta}{2} \cosh^{-1-2i\rho,l_0}\frac{\tau}{2}
\sum^{l_0}_{\dot{k}=-l_0}i^{\pm 1-\dot{k}} \tan^{\pm
1-\dot{k}}\frac{\theta}{2}
\tanh^{\dot{n}-\dot{k}}\frac{\tau}{2}\times\\
\hypergeom{2}{1}{\pm 1-l_0+1,1-l_0-\dot{k}} {\pm 1-\dot{k}+1}
{i^2\tan^2\frac{\theta}{2}}
\hypergeom{2}{1}{\dot{n}+i\rho+\frac{1}{2},\frac{1}{2}+i\rho-\dot{k}}
{\dot{n}-\dot{k}+1}{\tanh^2\frac{\tau}{2}}\nonumber
\end{multline}
are associated hyperspherical functions defined on the spheres
$S^2_c$ and $\dot{S}^2_c$. General solutions are found via the
following generalized Fourier integrals:
\[
\boldsymbol{\phi}_{\pm 1}(\balpha)= \lf
2(2\pi)^3\rf^{-\frac{1}{2}}\sum^\infty_{l_0=1}\boldsymbol{f}^{-\frac{1}{2}+i\rho,l_0}_{1,\pm
1}(r)\sum^{l_0}_{n=-l_0}\int\limits_{T_4}
\ar\begin{pmatrix}\varepsilon_\pm(\bk)\\
\varepsilon_\pm(\bk)\end{pmatrix} e^{ikx} \alpha^{\pm
1}_{l_0,n}\fM_{\pm
1,n}^{-\frac{1}{2}+i\rho,l_0}(\varphi,\epsilon,\theta,\tau,0,0)d^4x,
\]
\[
\dot{\boldsymbol{\phi}}_{\pm 1}(\balpha)= \lf
2(2\pi)^3\rf^{-\frac{1}{2}}\sum^\infty_{l_0=1}\boldsymbol{f}^{-\frac{1}{2}-i\rho,\dot{l}_0}_{1,\pm
1}(r^\ast) \sum^{l_0}_{\dot{n}=-l_0}\int\limits_{T_4}
\ar\begin{pmatrix}\varepsilon^\ast_\pm(\bk)\\
\varepsilon^\ast_\pm(\bk)\end{pmatrix} e^{-ikx}\alpha^{\pm
1}_{l_0,\dot{n}} \fM_{\pm
1,\dot{n}}^{-\frac{1}{2}-i\rho,l_0}(\varphi,\epsilon,\theta,\tau,0,0)d^4x,
\]
where
\begin{eqnarray}
\alpha^{\pm 1}_{l_0,n}&=&\frac{(-1)^n(2l_0+1)^2}{32\pi^4
\boldsymbol{f}^{-\frac{1}{2}+i\rho,l_0}_{1,\pm 1}(a)}
\int\limits_{S^2_c}\int\limits_{T_4}F_{\pm 1}(\balpha)e^{-ikx}
\fM^{-\frac{1}{2}+i\rho,l_0}_{\pm 1,n}(\varphi,\epsilon,\theta,\tau,0,0)d^4xd^4\fg,\nonumber\\
\alpha^{\pm 1}_{l_0,\dot{n}}&=&
\frac{(-1)^{\dot{n}}(2l_0+1)^2}{32\pi^4
\boldsymbol{f}^{-\frac{1}{2}-i\rho,\dot{l}_0}_{1,\pm 1}(a^\ast)}
\int\limits_{S^2_c}\int\limits_{T_4} \dot{F}_{\pm 1}(\balpha)e^{ikx}
\fM^{-\frac{1}{2}-i\rho,l_0}_{\pm 1,\dot{n}}
(\varphi,\epsilon,\theta,\tau,0,0)d^4xd^4\fg.\nonumber
\end{eqnarray}
\section{Interaction}
Up to now we analyze free Dirac and Maxwell fields. Let us consider
an interaction between these fields. As usual, interactions between
the fields are described by an interaction Lagrangian $\cL_I$. In
our case we take the following Lagrangian \cite{Var06b}:
\begin{equation}\label{LagInt}
\cL_I(\balpha)=\mu(\overline{\boldsymbol{\psi}}(\balpha)
\sigma^D_{(\mu\nu)^k}\boldsymbol{\psi}(\balpha))
(\Xi^M_{(\rho)k}\boldsymbol{\phi}(\balpha)),
\end{equation}
where
$\sigma^D_{\mu\nu}=\frac{1}{2}(\Xi^D_\mu\Xi^D_\nu-\Xi^D_\nu\Xi^D_\mu)$
and
$\Xi^D=(\Gamma^D_0,\Gamma^D_1,\Gamma^D_2,\Gamma^D_3,\Upsilon^D_1,
\Upsilon^D_2,\Upsilon^D_3,\Upsilon^D_4,\Upsilon^D_5,\Upsilon^D_6)$,
$\Xi^M=(\Gamma^M_0,\Gamma^M_1,\Gamma^M_2,\Gamma^M_3,\Upsilon^M_1,\Upsilon^M_2,
\Upsilon^M_3,\Upsilon^M_4,\Upsilon^M_5,\Upsilon^M_6)$, here
$\Gamma^D$ and $\Upsilon^D$ are the matrices (\ref{Gamma1}) and
(\ref{Upsilon1})--(\ref{Upsilon2}), and $\Gamma^M$ and $\Upsilon^M$
are the matrices (\ref{Gamma2}) and
(\ref{Upsilon3})--(\ref{Upsilon4}).

The full Lagrangian of interacting Dirac and Maxwell fields equals
to a sum of the free field Lagrangians and the interaction
Lagrangian:
\[
\cL(\balpha)=\cL_D(\balpha)+\cL_M(\balpha)+\cL_I(\balpha),
\]
where $\cL_D(\balpha)$ and $\cL_M(\balpha)$ are of the type
(\ref{LagDir}) and (\ref{LagMax}), respectively. Or,
\begin{multline}
\cL(\balpha)=-\frac{1}{2}\left(\overline{\boldsymbol{\psi}}(\balpha)
\Xi^D_\mu\frac{\partial\boldsymbol{\psi}(\balpha)}{\partial\balpha_\mu}
-\frac{\partial\overline{\boldsymbol{\psi}}(\balpha)}
{\partial\balpha_\mu}\Xi^D_\mu\boldsymbol{\psi}(\balpha)\right)-\\
-\frac{1}{2}\left(\overline{\boldsymbol{\phi}}(\balpha)
\Xi^M_\mu\frac{\partial\boldsymbol{\phi}(\balpha)}{\partial\balpha_\mu}
-\frac{\partial\overline{\boldsymbol{\phi}}(\balpha)}
{\partial\balpha_\mu}\Xi^M_\mu\boldsymbol{\phi}(\balpha)\right)-\\
-\kappa\overline{\boldsymbol{\psi}}(\balpha)\boldsymbol{\psi}(\balpha)+
\mu(\overline{\boldsymbol{\psi}}(\balpha)
\sigma^D_{(\mu\nu)^k}\boldsymbol{\psi}(\balpha))
(\Xi^M_{(\rho)k}\boldsymbol{\phi}(\balpha)).\nonumber
\end{multline}
Since the Lagrangian (\ref{LagInt}) does not contain derivatives on
the field functions, then for a Hamiltonian density we have
$\cH_I(\balpha)=-\cL_I(\balpha)$.

As is known, in the standard quantum field theory the $S$-matrix is
expressed via the Dyson formula \cite{Sch61}
\begin{equation}\label{Dyson}
S=T\left[\exp\left(-\frac{i}{\hbar c}\int\limits^{+\infty}_{-\infty}
\cH_I(x)d^4x\right)\right],
\end{equation}
where $T$ is the time ordering operator.

In our case, the electron-positron and photon fields are defined on
the space $\cM_8=\R^{1,3}\times S^2_c$ which larger then the
Minkowski space $\R^{1,3}$. With a view to define a formula similar
to the equation (\ref{Dyson}) it is necessarily to replace $d^4x$ by
the following invariant measure on $\cM_8$:
\[
d^8\mu=d^4xd^4\fg,
\]
where
\[
d^4\fg=\sin\theta^cd\theta d\tau d\varphi d\epsilon.
\]
Therefore, an analogue of the Dyson formula (\ref{Dyson}) on the
manifold $\cM_8$ can be written as follows
\[
S=T\left[\exp\left(-\frac{i}{\hbar c}
\int\limits_{T_4}\int\limits_{S^2_c}
\cH_I(\balpha)d^4xd^4\fg\right)\right].
\]

Let us consider an integral on trilinear form, defining the
interaction, over the two-dimensional complex sphere:
\[
\int\limits_{S^2_c}\cH_I(\balpha)dS^2_c\sim\int\limits_{S^2_c}\fM^{\dot{m}_e}_{\dot{l}_e}
(\varphi,\epsilon,\theta,\tau,0,0)\fM^{m_e}_{l_e}
(\varphi,\epsilon,\theta,\tau,0,0)\fM^{m_f}_{-\frac{1}{2}+i\rho}
(\varphi,\epsilon,\theta,\tau,0,0)\sin\theta^cd\theta^cd\varphi^c,
\]
that is, to investigate convergence of the integral
\[
I_1=\int\limits^{+\infty}_{-\infty}\int\limits^{+\infty}_{-\infty}e^{-i(
m_e\varphi^c-\dot{m}_e\dot{\varphi}^c-m_f\varphi^c)}\fZ^{\dot{m}_e}_{\dot{l}_e}(\cos\dot{\theta}^c)
\fZ^{m_e}_{l_e}(\cos\theta^c)\fZ^{m_f}_{-\frac{1}{2}+i\rho}(\cos\theta^c)\sin\theta^cd\theta^cd\varphi^c.
\]
Here the symbols $l_e$, $m_e$ correspond to the electron field
$\boldsymbol{\psi}(\fg)$, $\dot{l}_e$, $\dot{m}_e$ correspond to the
positron field $\overline{\boldsymbol{\psi}(\fg)}$, and $m_f$
corresponds to the photon field $\boldsymbol{\phi}(\fg)$. It is
obvious that convergence of the integral
\[
I_2=\int\limits^{+\infty}_{-\infty}e^{-i(
m_e\varphi^c-\dot{m}_e\dot{\varphi}^c-m_f\varphi^c)}d\varphi^c
\]
is not difficult to investigate. Let us consider the following
integral:
\[
I_3=\int\limits^{+\infty}_{-\infty}\fZ^{\dot{m}_e}_{\dot{l}_e}(\cos\dot{\theta}^c)
\fZ^{m_e}_{l_e}(\cos\theta^c)\fZ^{m_f}_{-\frac{1}{2}+i\rho}(\cos\theta^c)\sin\theta^cd\theta^c.
\]
Rewriting hyperspherical functions via the hypergeometric functions,
we obtain
\begin{multline}
I_3=\frac{i^{m_e+\dot{m}_e+m_f}}{\Gamma(m_e+1)\Gamma(\dot{m}_e+1)\Gamma(m_f+1)}
\sqrt{\frac{\Gamma(l_e+m_e+1)\Gamma(\dot{l}_e+\dot{m}_e+1)\Gamma(m_f+i\rho+\frac{1}{2})}
{\Gamma(l_e-m_e+1)\Gamma(\dot{l}_e-\dot{m}_e+1)\Gamma(i\rho-m_f+\frac{1}{2})}}\times\\
\int\limits^{+\infty}_{-\infty}\sin^{\dot{m}_e+m_e+m_f}\frac{\theta^c}{2}
\cos^{\dot{m}_e+m_e+m_f}\frac{\theta^c}{2}
\hypergeom{2}{1}{\dot{l}_e+\dot{m}_e+1,\dot{m}_e-\dot{l}_e}{\dot{m}_e+1}{\sin^2\frac{\theta^c}{2}}\times\\
\hypergeom{2}{1}{l_e+m_e+1,m_e-l_e}{m_e+1}{\sin^2\frac{\theta^c}{2}}
\hypergeom{2}{1}{i\rho+m_f+\frac{1}{2},m_f-i\rho
+\frac{1}{2}}{m_f+1}{\sin^2\frac{\theta^c}{2}}\sin\theta^cd\theta^c.
\label{Integ3}
\end{multline}
Further, using an explicit expression for the associated
hyperspherical function
\begin{multline}
\fZ^m_l(\cos\theta^c)=i^m\sqrt{\frac{\Gamma(l-m+1)}{\Gamma(l+m+1)}}\cos^m\frac{\theta^c}{2}
\sin^m\frac{\theta^c}{2}\times\\
\sum^{l-m}_{j=0}\frac{(-1)^j\Gamma(l+m+j+1)}{\Gamma(j+1)\Gamma(m+j+1)\Gamma(l-m-j+1)}\sin^{2j}\frac{\theta^c}{2},
\nonumber
\end{multline}
we rewrite the first two hypergeometric functions in the integral
(\ref{Integ3}). Then
\begin{multline}
I_3=\frac{i^{m_e+\dot{m}_e+m_f}}{\Gamma(m_f+1)}
\sqrt{\frac{\Gamma(l_e-m_e+1)\Gamma(\dot{l}_e-\dot{m}_e+1)\Gamma(m_f+i\rho+\frac{1}{2})}
{\Gamma(l_e+m_e+1)\Gamma(\dot{l}_e+\dot{m}_e+1)\Gamma(i\rho-m_f+\frac{1}{2})}}\times\\
\sum^{l_e-m_e}_{j_1=0}\sum^{\dot{l}_e-\dot{m}_e}_{j_2=0}\frac{(-1)^{j_1+j_2}\Gamma(l_e+m_e+j_1+1)
}{\Gamma(j_1+1)\Gamma(j_2+1)
\Gamma(m_e+j_1+1)\Gamma(\dot{m}_e+j_2+1)}\times\\
\int\limits^{+\infty}_{-\infty}\cos^{m_e+\dot{m}_e+m_f}\frac{\theta^c}{2}
\sin^{m_e+\dot{m}_e+m_f+2j_1+2j_2}\frac{\theta^c}{2}\times\\
\hypergeom{2}{1}{i\rho+m_f+\frac{1}{2},m_f-i\rho
+\frac{1}{2}}{m_f+1}{\sin^2\frac{\theta^c}{2}}\sin\theta^cd\theta^c.
\nonumber
\end{multline}
Making the substitution $z=\cos\theta^c$ in the integral
\begin{multline}
I_4=\frac{1}{2^{m_e+\dot{m}_e+m_f}}\int\limits^{+\infty}_{-\infty}\sin^{m_e+\dot{m}_e+m_f}\theta^c
\sin^{2j_1+2j_2}\frac{\theta^c}{2}\times\\
\hypergeom{2}{1}{i\rho+m_f+\frac{1}{2},m_f-i\rho
+\frac{1}{2}}{m_f+1}{\sin^2\frac{\theta^c}{2}}\sin\theta^cd\theta^c,
\nonumber
\end{multline}
we obtain
\begin{multline}
I_4=\frac{1}{2^{m_e+\dot{m}_e+m_f+j_1+j_2}}\int\limits^{+\infty}_{-\infty}
(1-z^2)^{\frac{m_e+\dot{m}_e+m_f}{2}}\left(\frac{1-z}{2}\right)^{j_1+j_2}\times\\
\hypergeom{2}{1}{i\rho+m_f+\frac{1}{2},m_f-i\rho
+\frac{1}{2}}{m_f+1}{\frac{1-z}{2}}dz. \nonumber
\end{multline}
Or,
\begin{multline}
I_4=\frac{1}{2^{m_e+\dot{m}_e+m_f+j_1+j_2}}\sum^{m_e+\dot{m}_e+m_f}_{q=0}(-1)^q
\frac{( m_e+\dot{m}_e+m_f)!}{q!(m_e+\dot{m}_e+m_f-q)!}\times\\
\int\limits^{+\infty}_{-\infty}
z^k\left(\frac{1-z}{2}\right)^{j_1+j_2}
\hypergeom{2}{1}{i\rho+m_f+\frac{1}{2},m_f-i\rho
+\frac{1}{2}}{m_f+1}{\frac{1-z}{2}}dz. \nonumber
\end{multline}
Introducing a new variable $t=(1-z)/2$, we find
\begin{multline}
I_4=\frac{1}{2^{m_e+\dot{m}_e+m_f+j_1+j_2}}\sum^{m_e+\dot{m}_e+m_f}_{q=0}(-1)^{q+1}
\frac{( m_e+\dot{m}_e+m_f)!}{q!(m_e+\dot{m}_e+m_f-q)!}\times\\
\int\limits^{+\infty}_{-\infty}(1-2t)^qt^{j_1+j_2}\hypergeom{2}{1}{i\rho+m_f+\frac{1}{2},m_f-i\rho
+\frac{1}{2}}{m_f+1}{t}dt. \nonumber
\end{multline}
Decomposing $(1-2t)^q$ via the Newton binomial, we obtain
\begin{multline}
I_4=\frac{1}{2^{m_e+\dot{m}_e+m_f+j_1+j_2}}\sum^{m_e+\dot{m}_e+m_f}_{q=0}
\sum^q_{p=0}(-1)^{q+p+1}
\frac{( m_e+\dot{m}_e+m_f)!}{(m_e+\dot{m}_e+m_f-q)!}\times\\
\frac{2^p}{p!(k-p)!}\int\limits^{+\infty}_{-\infty}t^{j_1+j_2+p}\hypergeom{2}{1}{i\rho+m_f+\frac{1}{2},m_f-i\rho
+\frac{1}{2}}{m_f+1}{t}dt. \nonumber
\end{multline}
With the aim to calculate the latter integral we use the following
formula \cite{Prud}:
\begin{multline}
I_5=\int t^n\hypergeom{2}{1}{a,b}{c}{t}dt=\\
=n!\sum^{n+1}_{k=1}(-1)^{k+1}\frac{(c-k)_kt^{n-k+1}}{(a-k+1)!(a-k)_k(b-k)_k}
\hypergeom{2}{1}{a-k,b-k}{c-k}{t}.\nonumber
\end{multline}
Then
\begin{multline}
I_4=\frac{1}{2^{m_e+\dot{m}_e+m_f+j_1+j_2}}\sum^{m_e+\dot{m}_e+m_f}_{q=0}
\sum^q_{p=0}(j_1+j_2+p)!\times\\
\sum^{j_1+j_2+p}_{k=1}
\frac{(-1)^{q+p+k}2^p(m_e+\dot{m}_e+m_f)!}{p!(k-p)!(m_e+\dot{m}_e+m_f-q)!(m_f+i\rho+\frac{3}{2}-k)!}\times\\
\frac{(m_f-k+1)_kt^{j_1+j_2+p-k+1}}{(m_f+i\rho-k+\frac{1}{2})_k(m_f-k+1)_k}\times\\
\hypergeom{2}{1}{i\rho+m_f-k+\frac{1}{2},m_f-i\rho-k
+\frac{1}{2}}{m_f-k+1}{t}dt. \nonumber
\end{multline}
With the aim to investigate the convergence of (\ref{Integ3}) let us
apply the following asymptotic expansion for the hypergeometric
function \cite{Bat}:
\begin{multline}
\hypergeom{2}{1}{a,b}{c}{t}=\frac{\Gamma(c)\Gamma(b-a)}{\Gamma(b)
\Gamma(c-a)}(-t)^{-a}
\hypergeom{2}{1}{a,1-c+a}{1-b+a}{\frac{1}{t}}+\\
+\frac{\Gamma(c)\Gamma(a-b)}{\Gamma(a)\Gamma(c-b)}(-t)^{-b}
\hypergeom{2}{1}{b,1-c+b}{1-a+b}{\frac{1}{t}}. \nonumber
\end{multline}
Thus,
\begin{multline}
I_3=\frac{i^{m_e+\dot{m}_e+m_f}}{\Gamma(m_f+1)}\sqrt{\frac{\Gamma(l_e-m_e+1)\Gamma(\dot{l}_e-\dot{m}_e+1)
\Gamma(m_f+i\rho+\frac{1}{2})}{\Gamma(l_e+m_e+1)\Gamma(\dot{l}_e+\dot{m}_e+1)\Gamma(i\rho-m_f+\frac{1}{2})}}\times\\
\sum^{l_e-m_e}_{j_1=0}\sum^{\dot{l}_e-\dot{m}_e}_{j_2=0}\frac{(-1)^{j_1+j_2}\Gamma(l_e+m_e+j_1+1)
}{2^{m_e+\dot{m}_e+m_f+j_1+j_2}\Gamma(j_1+1)\Gamma(j_2+1)
\Gamma(m_e+j_1+1)\Gamma(\dot{m}_e+j_2+1)}\times\nonumber
\end{multline}
\begin{multline}
\frac{\Gamma(\dot{l}_e+\dot{m}_e+j_2+1)}{\Gamma(l_e-m_e-j_1+1)\Gamma(\dot{l}_e-\dot{m}_e-j_2+1)}\times\\
\sum^{m_e+\dot{m}_e+m_f}_{q=0}\sum^q_{p=0}(j_1+j_2+p)!\sum^{j_1+j_2+p}_{k=1}
\frac{(-1)^{q+p+k}2^p(m_e+\dot{m}_e+m_f)!}{p!(k-p)!(m_e+\dot{m}_e+m_f-q)!(m_f+i\rho+\frac{3}{2}-k)!}\times\nonumber
\end{multline}
\begin{multline}
\frac{(m_f-k+1)_k}{(m_f+i\rho-k+\frac{1}{2})_k(m_f-k+1)_k}
\left[(-1)^{k-m_f-i\rho-\frac{1}{2}}\frac{\Gamma(m_f-k+1)\Gamma(-2i\rho)}
{\Gamma(m_f-i\rho-k+\frac{1}{2})\Gamma(\frac{1}{2}-i\rho)}\times\right.\\
t^{j_1+j_2+p-m_f-i\rho+\frac{1}{2}}
\hypergeom{2}{1}{m_f+i\rho-k+\frac{1}{2},i\rho+\frac{1}{2}}{2i\rho+1}{\frac{1}{t}}+
\nonumber
\end{multline}
\begin{multline}
\left.(-1)^{k-m_f+i\rho-\frac{1}{2}}\frac{\Gamma(m_f-k+1)\Gamma(2i\rho)}
{\Gamma(m_f+i\rho-k+\frac{1}{2})\Gamma(\frac{1}{2}-i\rho)}\times\right.\\
\left.t^{j_1+j_2+p-m_f+i\rho+\frac{1}{2}}
\hypergeom{2}{1}{m_f+i\rho-k+\frac{1}{2},\frac{1}{2}-i\rho}{1-2i\rho}{\frac{1}{t}}\right],
\nonumber
\end{multline}
where $t=\sin^2\frac{\theta^c}{2}$. Since $I_3\sim 1/t^M$ (the
hypergeometric function ${}_2F_1$ can be written as a power series
in $1/t$) and $M\rightarrow\infty$, then $I_3$ converges at
$M>j_1+j_2+p-m_f+i\rho+\frac{1}{2}$ and, therefore, in accordance
with (\ref{Dyson}) the elements of $S$-matrix are defined by
convergent expressions.

\section{Summary}
We have proved convergence of quantum electrodynamics on the
Poincar\'{e} group in the case of homogeneous space
$\cM_8=\R^{1,3}\times S^2_c$. We considered Dirac like equations for
extended objects on $\cM_8$ and their particular solutions
corresponding to the fields of spin 1/2 and 1. At this point, the
spin-1 field (Maxwell field) is defined within an
infinite-dimensional representation of the Lorentz group. We showed
that an analogue of the Dyson formula for $S$-matrix in the case of
$\cM_8$ is defined by convergent integrals. It would be interesting
to consider quantum field models and their convergence on other
homogeneous spaces of the Poincar\'{e} group, such as
$\cM_6=\R^{1,3}\times S^2$, $\cM_7=\R^{1,3}\times H^3$ and
$\cM_{10}=\R^{1,3}\times\fL_6$. It would be interesting also to
consider wave equations and field models for the fields
$\boldsymbol{\psi}(\balpha)=\langle
x,\fq\,|\boldsymbol{\psi}\rangle$ on the de Sitter group, where
$x\in T_5$ and $\fq\in\spin_+(1,4)\simeq\Sp(1,1)$, and for the
fields $\boldsymbol{\psi}(\balpha)=\langle
x,\fc\,|\boldsymbol{\psi}\rangle$ on the conformal group, where
$x\in T_6$ and $\fc\in\spin_+(2,6)\simeq\SU(2,2)$. Our next paper
will be devoted to these questions.

\section*{Appendix: Spinor groups and bivector spaces}
As is known, rotations of pseudo-Euclidean spaces $\R^{p,q}$ are
defined by spinor groups \cite{Lips}
\[
\spin(p,q)=\left\{s\in\Lip^+_{p,q}\;|\;N(s)=\pm 1\right\},
\]
where $\Lip^+_{p,q}=\Lip_{p,q}\cap\cl^+_{p,q}$ is a special
Lipschitz group, $s\in\cl^+_{p,q}$ is an even invertible element of
the real Clifford algebra $\cl_{p,q}$. In more detail, the element
$s$ is a linear combination of even basic elements, that is,
\[
s=\sum_ka^{i_1\ldots i_{2k}}\e_{i_1\ldots i_{2k}}.
\]
The condition $N(s)=\pm 1$ means that
\begin{equation}\label{5.99}
\sum_k\sigma(i_1)\cdots\sigma(i_{2k})\left(a^{i_1\ldots
i_{2k}}\right)^2=\pm 1.
\end{equation}
On the other hand, from the fact that the elements
$s\in\Lip^+_{p,q}$ are even products of $\nu=\sum\nu^i\e_i$, it is
easy to derive that coordinates of $s\in\spin(p,q)$ are related by
the following conditions \cite{Roz55}:
\begin{equation}\label{5.100}
a^{i_1i_2\ldots i_{2k}}(a)^{k-1}=(2k-1)!!a^{[i_1i_2}a^{i_2i_3}\cdots
a^{i_{2k-1}i_{2k}]}.
\end{equation}
Conditions (\ref{5.100}) can be rewritten in the form
\begin{equation}\label{5.101}
\left.\begin{array}{rcl}
aa^{i_1i_2i_3i_4}&=&3!!a^{[i_1i_2}a^{i_3i_4]},\\
aa^{i_1i_2i_3i_4i_5i_6}&=&5!!a^{[i_1i_2}a^{i_3i_4i_5i_6]},\\
\hdotsfor[2]{3}\\
aa^{i_1i_2\ldots i_{2k}}&=&(2k-1)!!a^{[i_1i_2\ldots}a^{i_3i_4\ldots
i_{2k-1}i_{2k}]}.
\end{array}\right\}
\end{equation}
Conditions (\ref{5.99}) and (\ref{5.100}) express all coordinates of
the elements $s$, belonging to the spinor group $\spin(p,q)$, via
$n(n-1)/2$ coordinates $a^{ij}$, where $n=p+q$. The number of the
coordinates $a^{ij}$ coincides with the number of parameters of the
rotation group of the space $\R^{p,q}$. This fact shows that
expressions (\ref{5.99}) and (\ref{5.100}) form a full system of the
conditions separating the spinor group $\spin(p,q)$ from the algebra
$\cl_{p,q}$.

Further, parameters of the rotation group of the space $\R^{p,q}$
form a {\it bivector space} $\R^N$, where $N=\frac{n(n-1)}{2}$.
Indeed, Let $\R^{p,q}$ be the $n$-dimensional pseudo-Euclidean
space, $p+q=n$. Let us evolve in $\R^{p,q}$ all the tensors
satisfying the following two conditions: 1) a rank of the tensors is
even; 2) covariant and contravariant indexes are divided into
separate skewsymmetric pairs. Such tensors can be exemplified by
bivectors (skewsymmetric tensors of the second rank). The set of all
bivector tensor fields in $\R^{p,q}$ is called {\it a bivector set},
and its representation in a given point of $\R^{p,q}$ is called {\it
a local bivector set}. In any tensor from the bivector set we take
the each skewsymmetric pair $\alpha\beta$ as one collective index.
At this point, from the two possible pairs $\alpha\beta$ and
$\beta\alpha$ we fix only one, for example, $\alpha\beta$. The
number of all collective indexes is equal to $N=\frac{n(n-1)}{2}$.
In a given point, the bivector set of the space $\R^{p,q}$ with the
contravariant components defines in the collective indexes a vector
set, and the each vector of this set has $N$ components. Identifying
these vectors with the points of the $N$-dimensional manifold, we
come to an affine manifold $E^N$ if and only if this manifold admits
a Klein geometry with the group
\begin{gather}
\eta^{a^\prime}=A^{a^\prime}_a\eta^a,\quad
\eta^a=A^a_{a^\prime}\eta^{a^\prime},\nonumber\\
\det A^{a^\prime}_a\neq 0,\quad
A^a_{b^\prime}A^{b^\prime}_c=\delta^a_c,\nonumber
\end{gather}
where
\[
A^{a^\prime}_a\longrightarrow
A^{\alpha^\prime}_{[\alpha}A^{\beta^\prime}_{\beta]}.
\]
Thus, any local bivector set of the space $\R^{p,q}$ ($p+q=n$) can
be mapped onto the affine space $E^N$ . Therefore, $E^N$ is related
with the each point of the space $\R^{p,q}$. The space $E^N$ is
called {\it a bivector space}. It should be noted that the bivector
space is a particular case of the most general mathematical
construction called a Grassmannian manifold (a manifold of
$m$-dimensional planes of the affine space). In the case $m=2$ the
manifold of two-dimensional planes is isometric to the bivector
space, and the Grassmann coordinates in this case are called
Pluecker coordinates.

The metrization of the bivector space $E^N$ is given by the formula
(see \cite{Pet69})
\begin{equation}\label{Metric}
g_{ab}\longrightarrow g_{\alpha\beta\gamma\delta}\equiv
g_{\alpha\gamma}g_{\beta\delta}-g_{\alpha\delta}g_{\beta\gamma},
\end{equation}
where $g_{\alpha\beta}$ is a metric tensor of the space $\R^{p,q}$,
and the collective indexes are skewsymmetric pairs
$\alpha\beta\rightarrow a$, $\gamma\delta\rightarrow b$. After
introduction of $g_{ab}$, the bivector affine space $E^N$ is
transformed to a metric space $\R^N$.

As is known, any transformation from the rotation group of the space
$\R^{p,q}$ can be represented via $\frac{n(n-1)}{2}$ transformations
in the planes $(x_1,x_2)$, $(x_1,x_3)$, $\ldots$, $(x_{p+q},x_1)$.
The full number of the planes $(x_i,x_j)$ is equal to
$N=\frac{n(n-1)}{2}$, and the each plane $(x_i,x_j)$ corresponds to
the coordinate $a^{ij}$ of the spinor group $\spin(p,q)$.

\end{document}